\documentclass[
reprint,
superscriptaddress,
amsmath,amssymb,
aps,
floatfix,
]{revtex4-2}

\usepackage{graphicx}
\usepackage{placeins}
\usepackage{dcolumn}                    
\usepackage[utf8]{inputenc}
\usepackage{bm}
\usepackage{braket}
\usepackage{siunitx}
\usepackage{float}
\DeclareSIUnit{\belmilliwatt}{Bm}
\DeclareSIUnit{\dBm}{\deci\belmilliwatt}
\usepackage[T1]{fontenc}
\usepackage{hyperref}
\hypersetup{
    colorlinks=true,
    linkcolor=black,
    citecolor=black,
    filecolor=black,     
    urlcolor=black,
}
\usepackage{xr}
\usepackage{titlesec}
\titleformat{\section}
  {\bfseries\Large}{}{0em}{}
\titleformat{\subsection}
  {\bfseries}{}{0em}{}

\begin{document}

\title{Tracking spin qubit frequency variations over 912 days}
 
\author{Kenji Capannelli}
\author{Brennan Undseth}
\author{Irene Fern\'{a}ndez de Fuentes}
\author{Eline Raymenants}
\author{Florian~K.~Unseld}
\author{Oriol Pietx-Casas}
\author{Stephan~G.~J.~Philips}
\author{Mateusz~T.~M\k{a}dzik}
\affiliation{QuTech and Kavli Institute of Nanoscience, Delft University of Technology, Lorentzweg 1, 2628 CJ Delft, The Netherlands}
\author{Sergey~V.~Amitonov}
\author{Larysa Tryputen}
\affiliation{QuTech and Netherlands Organization for Applied Scientific Research (TNO), Stieltjesweg 1, 2628 CK Delft, The Netherlands}
\author{Giordano Scappucci}
\author{Lieven~M.~K.~Vandersypen}
\altaffiliation{Corresponding author: l.m.k.vandersypen@tudelft.nl}
\affiliation{QuTech and Kavli Institute of Nanoscience, Delft University of Technology, Lorentzweg 1, 2628 CJ Delft, The Netherlands}

%\date{\today}
 
\begin{abstract}
Solid-state qubits are sensitive to their microscopic environment, causing the qubit properties to fluctuate on a wide range of timescales. The sub-Hz end of the spectrum is usually dealt with by repeated background calibrations, which bring considerable overhead. It is thus important to characterize and understand the low-frequency variations of the relevant qubit characteristics. In this study, we investigate the stability of spin qubit frequencies in the Si/SiGe quantum dot platform. We find that the calibrated qubit frequencies of a six-qubit device vary by up to $\pm 100$ MHz while performing a variety of experiments over a span of 912 days. These variations are sensitive to the precise voltage settings of the gate electrodes, however when these are kept constant to within 15 $\mathrm{\mu}$V, the qubit frequencies vary by less than $\pm 7$ MHz over periods up to 36 days. During overnight scans, the qubit frequencies of ten qubits across two different devices show a standard deviation below 200 kHz within a 1-hour time window. The qubit frequency noise spectral density shows roughly a $1/f$ trend above $10^{-4}$ Hz and, strikingly, a steeper trend at even lower frequencies. 
\end{abstract}
 
\maketitle 

High-fidelity single~\cite{Yoneda2018, Lawrie23, Nakajima20, Zwerver22, Madzik22} and two-qubit gates~\cite{Madzik22, Noiri22, Xue2022, Mills22} have been demonstrated in semiconductor spin qubit platforms, motivating efforts towards large-scale quantum computing. Indeed, recent efforts have focused on increasing the size of quantum dot arrays~\cite{Borsoi24, Neyens24, Ha2025, Philips22, George2024, John2024}, but to retain universal control involves a large number of calibration parameters that typically out-scale the number of qubits. As a case in point, in Ref.~\cite{Philips22}, a calibration protocol of 108 different parameters was developed to demonstrate universal control of six qubits. Furthermore, these parameters were subject to slow variations over time, thus compromising the fidelity of state preparation, quantum gates and measurement. 

To overcome parameter variations in time, recent studies have aimed at efforts to implement feedback control or periodic recalibration to retain high fidelities~\cite{Nakajima20, Nakajima21, Philips22, Berritta2024physics, Berritta2024real}, similar to efforts with, for instance, superconducting qubits~\cite{Riste2012} or trapped ions~\cite{Park21}. For successful feedback control, recalibration must occur on a timescale similar or faster than the typical fluctuations of the calibrated parameters. In the semiconductor spin platform, for qubits encoded in the spin state of a single charge, one of the most important parameters to recalibrate periodically is the qubit frequency, because it impacts the calibration of many other parameters and empirically needs to be recalibrated more often than most other parameters~\cite{Philips22, Madzik22}. %Therefore, we focus our attention on the temporal variation of the qubit frequencies throughout this study.

Qubit frequency noise in semiconducting spin qubit platforms is typically attributed to two dominant sources: magnetic and charge noise. Magnetic noise arises predominantly from the nuclear spin bath, which couples to the qubit via the hyperfine interaction. For a dense bath of nuclear spins, the flip-flop rates between nuclear spins driven by the dipole-dipole interaction are high, causing a Brownian spin diffusion with a power spectral density $S_m(f)$ that scales as $1/f^2$~\cite{Burkard23}. This behavior has been observed in GaAs quantum dot qubits, where all atoms have a nuclear spin 3/2~\cite{Reilly08, Malinowski17}. Also in the silicon platform, nuclear spin noise can dominate dephasing despite the lower (4.7$\%$) abundance of spinful nuclei~\cite{Rojas24}. However, isotopically purifying the active layer to remove spinful $^{29}$Si isotopes has proved effective in greatly increasing coherence times~\cite{Veldhorst14}. The larger average distance between the nuclear spins leads to reduced flip-flop rates, which manifests in a power spectral density closer to a $1/f$-like scaling~\cite{Eng15,Madzik20,Burkard23}. 

Charge noise on the other hand couples to the qubit frequencies by perturbing the confinement potential of the quantum dot in combination with intrinsic or artificial spin-orbit coupling~\cite{Kawakami16, Hendrickx24}. This noise is typically attributed to an ensemble of two-level fluctuators (TLFs) which, under certain assumptions, will manifest in a power spectral density
$S_c(f)$ with a $1/f^\alpha$ dependence with $\alpha\approx1$ over a wide-frequency range~\cite{Dutta81, Gungordu19, Bermeister14, Mehmandoost24}. Experimental data typically captures this behavior, however in~\cite{Elsayed22}, a high-volume characterization of 231 charge noise measurements of Si/SiO$_2$ quantum dots across many different devices reports charge noise profiles with varying values of $\alpha\in[0,2]$ over a frequency range of approximately $[10^{-1},10^2]$ Hz. A characterization of charge noise in Si/SiGe over a much wider frequency range of approximately $[10^{-6},10^6]$ Hz reveals a charge noise spectrum with a fluctuating value of $\alpha$~\cite{Connors22}. All these observations are consistent with coupling to a low-density ensemble of TLFs~\cite{Mehmandoost24}. Depending on the distance of the TLF's to the qubits, charge noise experienced by nearby qubits can exhibit both temporal and spatial correlations ~\cite{Seedhouse23, Tanttu24, Yoneda23, Rojas23}. 

Despite the many studies investigating the origin and power spectral density of spin qubit frequencies, it remains important to investigate the variation of qubit frequencies over periods of months to years, a timescale relevant for future quantum processors deployed at a customer site~\cite{Eendebak20}. 

In this study, we examine the stability of the resonance qubit frequency in two devices; a six-dot linear array and a four-dot 2x2 array, both defined within a $^{28}$Si/SiGe heterostructure. In the linear device, the qubit frequency stability is analysed over a period of 912 days. For this analysis, we separate intrinsic qubit frequency variations in time from qubit frequency shifts resulting from changes in the gate voltages, the mixing chamber temperature, the applied magnetic field and thermal cycles. In both devices, Ramsey experiments are utilized to track qubit frequency variations over time periods ranging from 10-41 hours, repeated for a total of 46 times over a data collection period lasting several months. We also discuss the impact of qubit frequency variations on the fidelity of a standard resonantly driven 90 degree qubit rotation in the Bloch sphere. Finally, we perform Fourier analysis on the obtained datasets to characterize the qubit frequency noise spectral density in the range of approximately $[10^{-7},10^{-2}]$ Hz.

\section*{Results}

\begin{figure}[ht]
    \centering
    \includegraphics[width=\linewidth]{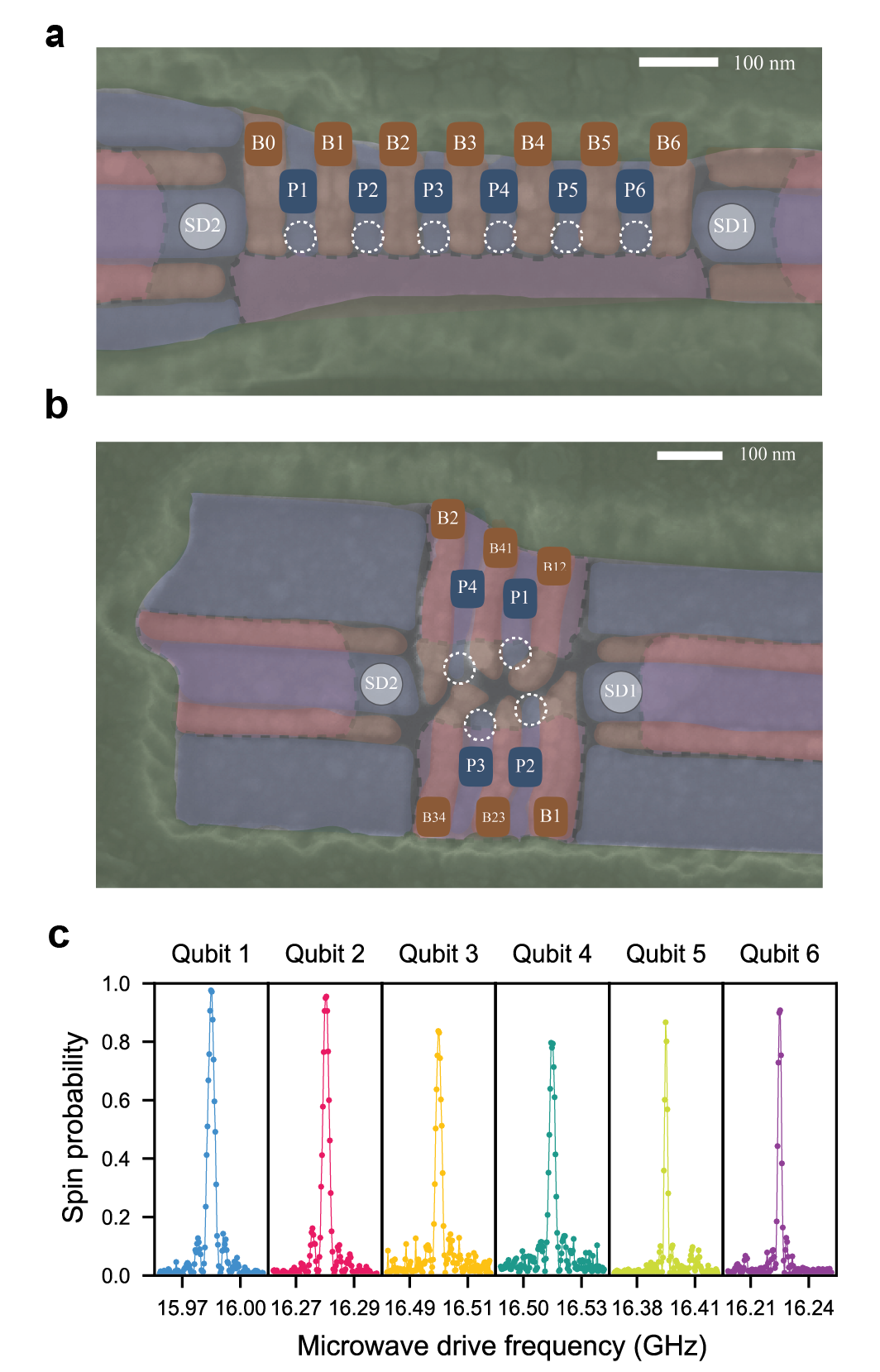}
    \caption{\textbf{Device layout and qubit frequency calibrations.} \textbf{a,b} False-colored scanning electron microscope images of two nominally identical devices to the ones used in the experiments. The linear device hosts six qubits, whilst the 2$\times$2 device hosts four qubits. In both devices, the electron spin qubits reside beneath the plunger gates (P) as indicated by the white dashed circles. The barrier gates (B) are used to tune the tunnel coupling between adjacent dots and to the reservoirs. Screening gates (colored pink and with their contours indicated by dashed black lines) are used for extra confinement of the dots. The screening gate located nearest to the dots is also used to deliver microwave signals for EDSR driving. \textbf{c} Example of resonance frequency calibration scans for all six qubits in the linear device taken on the same day.}
    \label{fig: devices}
\end{figure}

The two devices we study can host up to six and four qubits respectively, which we refer to as the linear~\cite{Philips22} and 2x2~\cite{Unseld23, Unseld24} devices. Nominally identical devices are depicted in Fig.~\ref{fig: devices}a and Fig.~\ref{fig: devices}b. Charges are isolated into quantum dots and an externally applied magnetic field induces a Zeeman splitting between the two spin states. Both devices employ two larger dots on the sides of the arrays for charge-sensing and utilize a Pauli spin blockade (PSB) protocol~\cite{Seedhouse21} to implement parity readout that is used for both initialization and readout of single-spin qubits~\cite{Loss1998}. The magnetized micromagnet induces a gradient magnetic field across the quantum dot arrays which allows for frequency-selective qubit addressing and for the manipulation of individual spin qubits via electric-dipole spin resonance (EDSR)~\cite{Pioro08}. The EDSR Hamiltonian for a single spin qubit in the rotating frame (setting $\hbar=1$ ) can be written in units of frequency as 
\begin{equation}\label{eq: EDSR rotating frame}
    H(t) = \frac{\Delta(t)}{2}\hat{\sigma}_z + \frac{\Omega(t)}{2}\Bigl(\cos(\Phi)\hat{\sigma}_x-\sin(\Phi)\hat{\sigma}_y\Bigr)
\end{equation}
where we define the qubit frequency detuning $\Delta(t) = f_0 - f_{MW} + \eta(t)$ as the difference between the qubit frequency at the time of calibration $f_0$ and the microwave drive frequency $f_{\rm{MW}}$, with the added presence of qubit frequency noise $\eta(t)$. Furthermore, $\Omega$ is the Rabi frequency (proportional to the microwave drive amplitude) and $\Phi$ sets the rotation axis in the rotating frame (which depends on the phase of the microwave drive). Finally, $\hat{\sigma}_i$ are the standard Pauli matrices. 

Qubit frequencies are calibrated by sweeping the frequency of the microwave burst to find the resonance condition as shown for the six qubits in the linear device in Fig.~\ref{fig: devices}c. These scans are fitted to the Rabi formula
\begin{equation}\label{eq: Rabi formula}
    P(\Delta) = \frac{\Omega^2}{\Omega^2+\Delta^2}\sin^2\Bigl(\pi t_{\rm{MW}}\sqrt{\Omega^2+\Delta^2}\Bigr)
\end{equation}
to extract the qubit frequency. Such calibration protocols are typically run on a daily basis. 

\subsection*{Qubit frequency calibrations over 912 days}

\begin{figure*}[t]
    \centering
    \includegraphics[width=\textwidth]{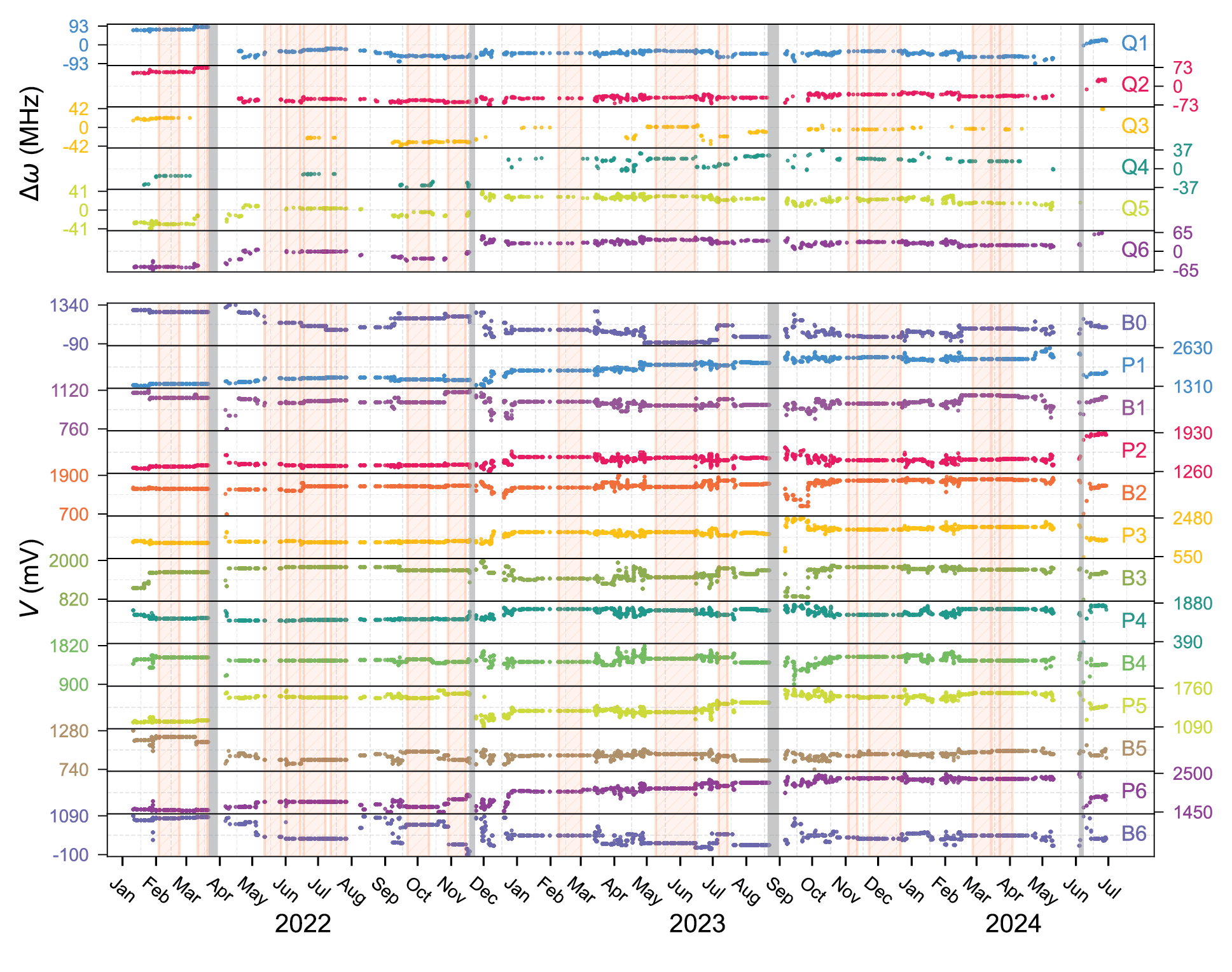}
    \caption{\textbf{Qubit frequency and DC voltage log}. A log of all six qubit frequencies (plotted in MHz deviations from the mean) and DC gate voltage settings in the linear device across a span of 912 days (from 24th Jan 2022 - 24th July 2024). The data points for the qubit frequencies are extracted from frequency scans as shown in Fig.~\ref{fig: devices}c (see main text). Some days feature multiple data points whereas for other days none are recorded. Any time a frequency data point is plotted for at least one of the qubits, the gate voltage settings for the entire array are included in the plot as well. The four gray regions indicate periods when the fridge underwent a thermal cycle. The 16 red regions indicate periods of longer than seven days when no gate voltage was changed by more than 15 $\mathrm{\mu}$V. The gates are labeled according to figure~\ref{fig: devices}a. The tickmarks on the horizontal axis are positioned in the middle of the month. This dataset forms the basis for further analysis in subsequent figures.}
    \label{fig: frequency log}
\end{figure*}

The linear device has been kept cold for several years, leading to multiple publications~\cite{Philips22, Undseth2023, DeFuentes25}. All data that has been collected, including calibration routines, are saved and indexed in a database, with each entry storing the experimental data and other relevant information (such as the dataset name, date and time) in an accompanying metadata file. This archive serves as a useful resource to extract long-term trends in device behavior. We analyzed qubit frequency calibration scans such as those shown in Fig.~\ref{fig: devices}c that were collected over a time span of 912 days (from 24/01/2022 - 24/07/2024). Each frequency scan was fitted to extract the qubit frequency and all corresponding gate voltage settings extracted from the experiment metadata file. The results are presented in Fig.~\ref{fig: frequency log}.

The fitting of the frequency scans was evaluated with a chi-square metric, which calculates the squared sum difference between the experimental data and the model, to filter out any low-quality, hence unreliable, scans (i.e. any scans that had poor spin contrast which made it difficult to reliably extract the qubit frequency). The plot only includes datapoints that passed this metric. For most of the 912 days, the magnetic field produced by the superconducting solenoid was set to 400 mT, but there were periods when the external field value was changed. Unfortunately, these changes were not logged. We therefore resort to filtering out any datapoints for which all six qubit frequencies were drastically different from their typical range. The fitting and filtering process is explained further in Supplementary Section~\ref{sup: filtering}.

The four gray regions indicate thermal cycles, whereby the sample was warmed up to room temperature and cooled down again without breaking vacuum of the dilution refrigerator. We find no significant impact of thermal cycles on the gate voltages required to find the qubits, apart from the initial tuning period immediately following a thermal cycle. In some thermal cycles, some of the qubit frequencies shift by up to several MHz, despite returning to the same nominal magnetic field value of 400 mT. To magnetize the micromagnet, the external field is first set to 2T before ramping down to 400 mT. Differences in this procedure could in part explain the small shifts in the observed qubit frequencies. The 16 red regions indicate periods longer than seven days where no DC gate voltage was changed by more than 15 $\mathrm{\mu}$V. These amount to a total of 259 days, which corresponds to the device operating in a practically stable voltage condition for 28\% of the 912 day period. Furthermore, during these stable periods, qubit frequencies would remain stable up to $\pm$ 7 MHz.
 %This is an order of magnitude larger than the typical range of Rabi frequencies achieved.
 
\begin{figure}[ht]
    \centering
    \includegraphics[width=\linewidth]{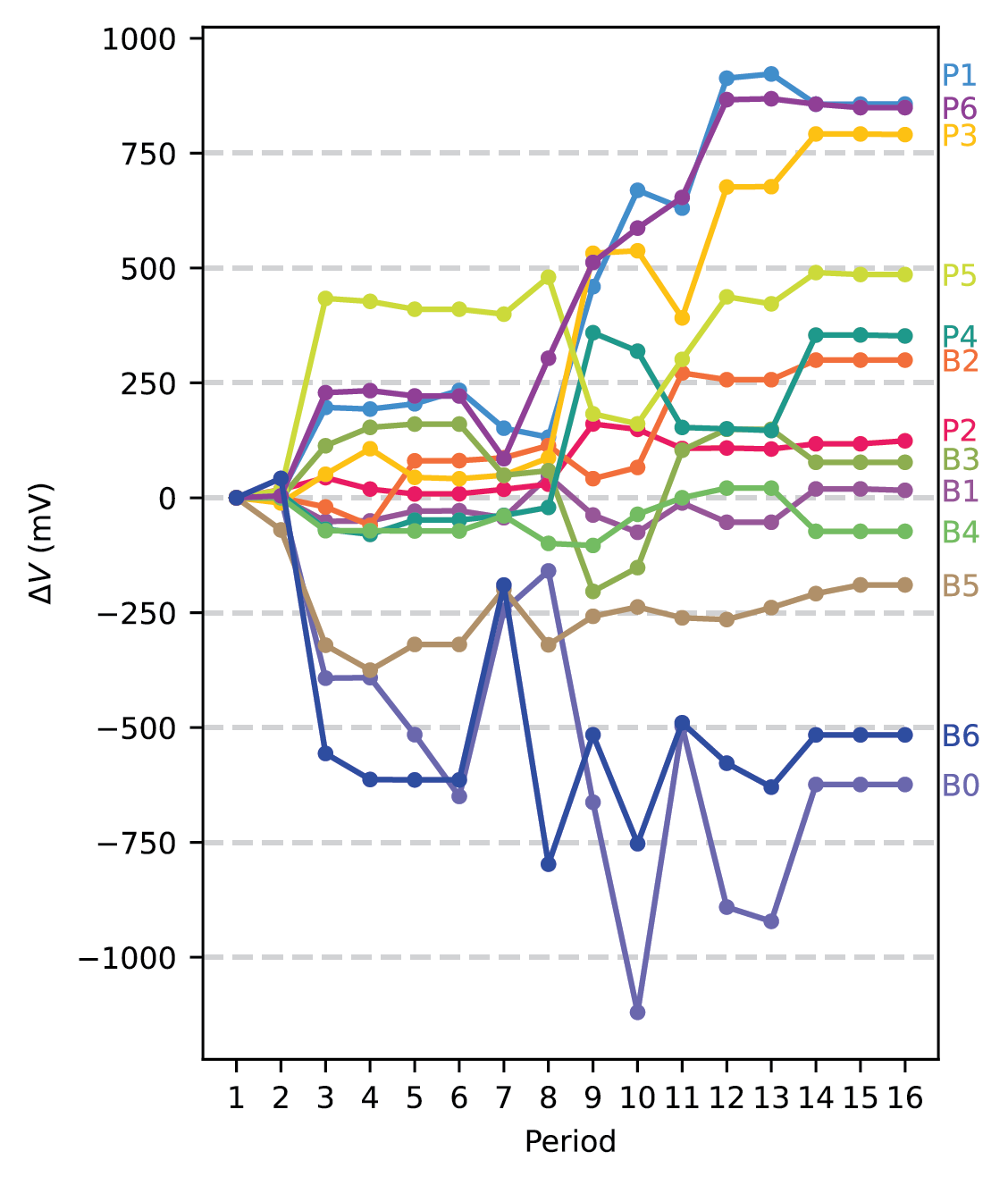}
    \caption{\textbf{Gate voltage drift over time.} Voltage difference $\Delta V = V_N - V_1$, where $V_1$ is the voltage setting corresponding to period 1, extracted from the 16 stable periods identified in Fig.~\ref{fig: frequency log} and displayed chronologically. This dataset highlights the drift in gate voltages required to operate the full 6-qubit array over the 912 day period. Most gate voltages do not drift by more than 500 mV.}
    \label{fig: gate voltage stability}
\end{figure}

\subsection*{Gate voltage stability}

Changes in the DC voltage settings cause the qubit frequency of nearby qubits to shift, either by modifying the $g$-factor~\cite{Liles21} or by displacing the spins in the stray magnetic field gradient induced by the micromagnet~\cite{Kobayashi24}. For example, between September and November 2023 in Fig.~\ref{fig: frequency log}, we notice that the qubit frequencies undergo an unstable period which coincides in time with a period of device tuning which requires many changes in the DC voltage settings. Even without targeted measurements to probe the sensitivity of the qubit frequencies to changes in individual gate voltages, a further analysis of the data in this segment of Fig.~\ref{fig: frequency log} already provides indications of correlations between qubit frequencies and gate voltages (see Supplementary Section~\ref{sup: q freq vs gate voltage}. Voltages are typically tuned to optimize tunnel couplings to the reservoir for loading and unloading of the dots and between adjacent dots for optimizing initialization, readout and two-qubit gate fidelities. Furthermore, voltages are sometimes retuned in an attempt to reduce the impact of strongly-coupled TLFs~\cite{Ye24A}.

\begin{figure}[ht]
    \centering
    \includegraphics[width=\linewidth]{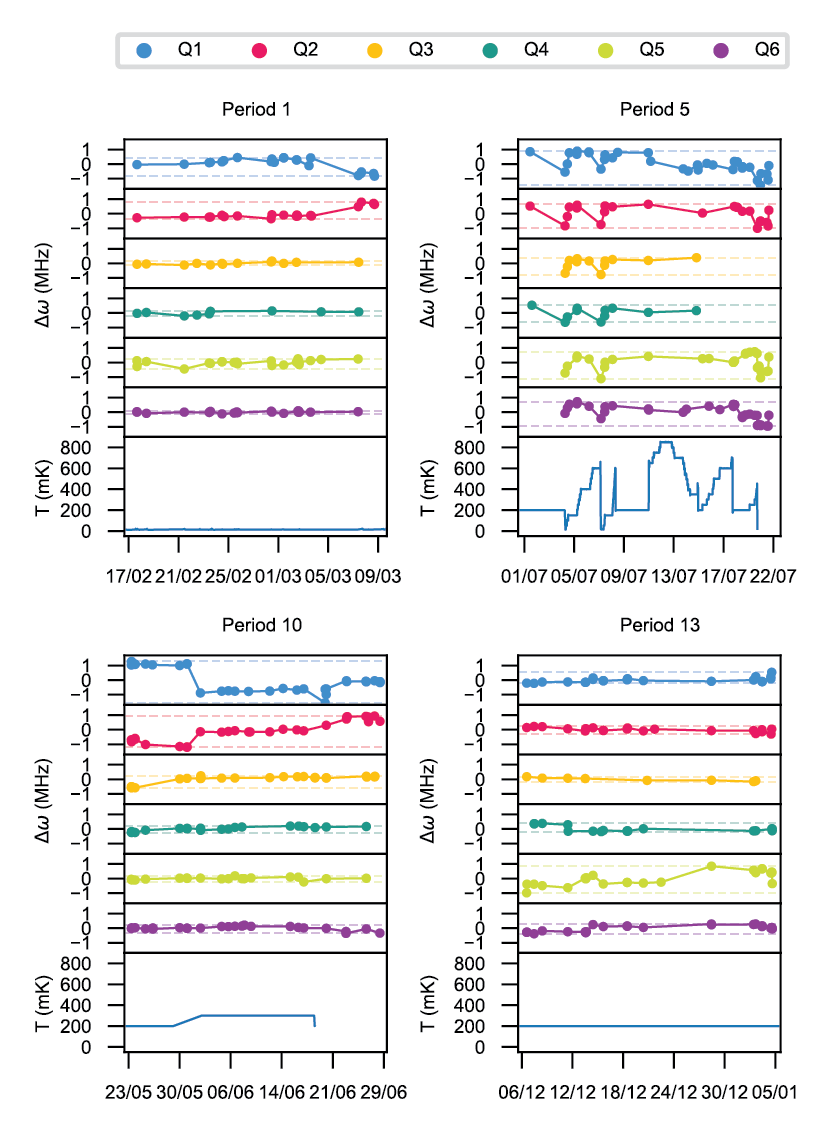}
    \caption{\textbf{Qubit frequency variations over stable voltage periods} (obtained from Fig.~\ref{fig: frequency log}). Each plot corresponds to the qubit frequency variations in the stable period indicated above the panel. In each panel, the top six subpanels correspond to qubit frequency variations, whilst the bottom subpanel shows the mixing chamber temperature at the time of data collection. \textbf{a} Data taken at base temperature of 12.8 mK. \textbf{b} Data taken while the mixing chamber temperature was varied from 12.8 mK to about 800 mK. We observe that qubit frequency shifts globally correlate with the mixing chamber temperature over this period. \textbf{c} Data taken at 200 mK before changing to 300 mK. The temperature was no longer logged after 18/06. \textbf{d} Data taken at 200 mK.}
    \label{fig: stable periods}
\end{figure}

We also characterized the long-term variations in gate voltages by tracking the set DC voltages for each stable period. In Fig.~\ref{fig: gate voltage stability}, we plot the difference in gate voltage setting from period 1 to 16. We observe that most set values did not change by more than 500 mV over the 912 day period, with the exception of gates P1, P3, P6, B0 and B6. Apart from P3, these gates are at the edges of the array and are used to control the loading/unloading of the dots. All plunger gate voltages undergo a net positive change whereas the majority of the barrier gate voltages undergo a negative change. The outer barrier gate voltages B0 and B6 see the most drastic negative change, which was intended to reduce the tunnel coupling to the sensing dots to operate the device in ``isolated'' mode. In order to keep the electron number in dots 1 and 6 fixed, the plungers P1 and P6 had to be taken to larger values.

\subsection*{Long-term qubit frequency variations extracted from frequency scans}

Altogether, including thermal cycles and gate voltage changes, and with a variety of experiments running, the qubit frequency shifts seen in Fig.~\ref{fig: frequency log} range from $\pm 37$ MHz for Q4 to $\pm 93$ MHz for Q1. Once a stable operating condition of magnetic field and DC gate voltages is established (the red regions in Fig.~\ref{fig: frequency log}), we observe an average range and standard deviation in the qubit frequency of 867 kHz and 268 kHz respectively, averaged over all qubits and stable periods. 

To get further insight on the variations of the qubit frequencies during the stable periods, we inspect the longest stable periods with consistent frequency calibrations in Fig.~\ref{fig: stable periods}, extending over 19, 20, 36 and 29 days respectively. We notice some variability in exactly how steady the frequency of each qubit is in any given stable period. Motivated by the findings in ~\cite{Undseth2023}, where a dependence of the qubit frequencies with temperature was reported, we includes plots of the mixing chamber temperature below the qubit frequency evolution for each period.
In Fig.~\ref{fig: stable periods}b, we can notice clear global correlations between all qubit frequencies, which in this case are also correlated with the temperature changes of the mixing chamber (from nominally 12.8 mK to about 800 mK).

\begin{figure}[ht]
    \centering
    \includegraphics[width=\linewidth]{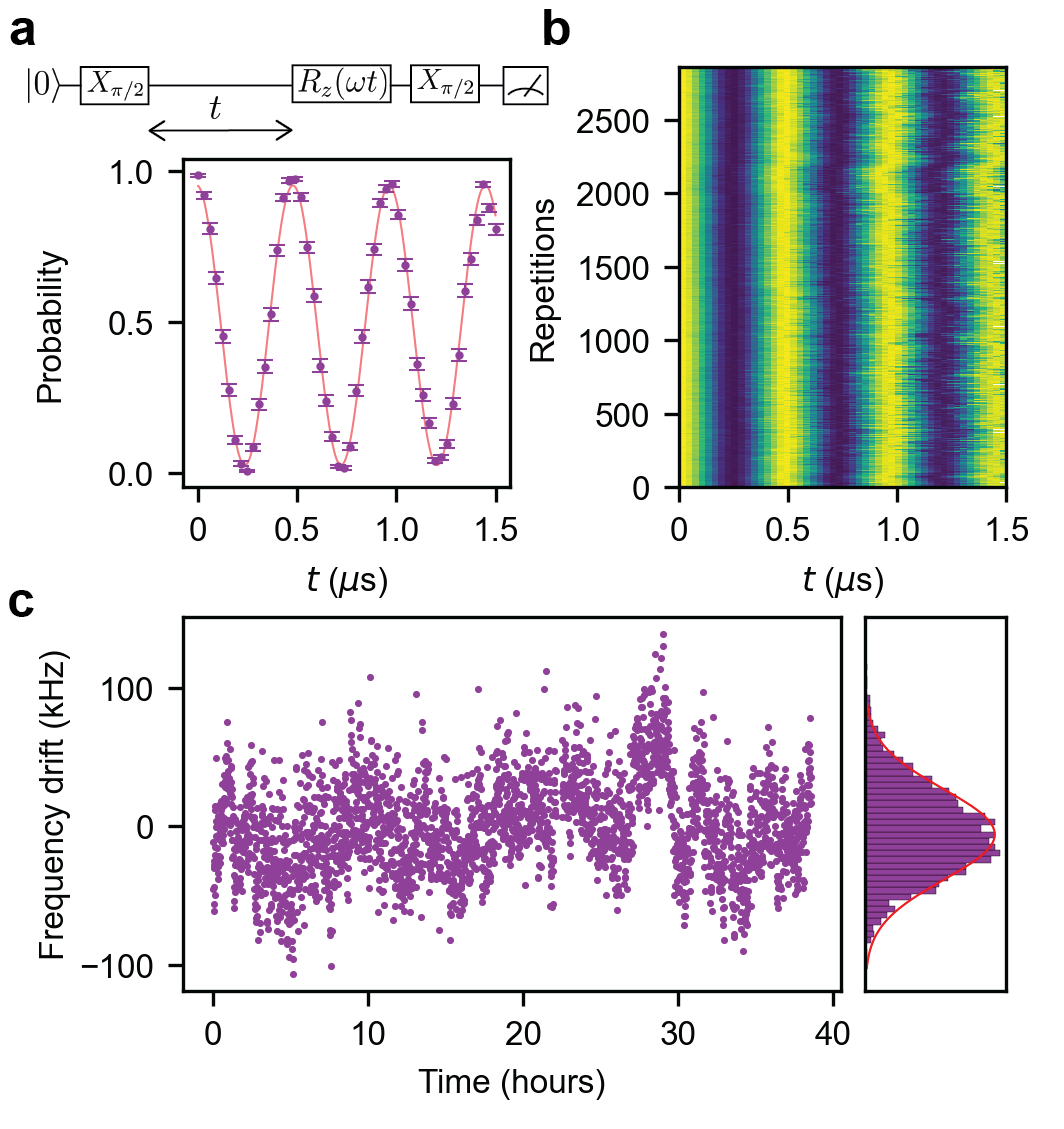}
    \caption{\textbf{Ramsey experiments to track qubit frequency variations.} \textbf{a} Circuit diagram of the Ramsey sequence used in the experiments along with an example measurement trace. Error bars correspond to the 95\% binomial confidence interval of the number of recorded spin up measurements with respect to the total number of shots and the red line is fit to Eq.~\ref{eq: Ramsey}. The microwave drive frequency is fixed to the pre-calibrated qubit frequency $f_{\mathrm{MW}}=f_0$. A virtual detuning of $\omega=2$ MHz is realized by a virtual rotation $R_z(\omega t)$ to enable a more reliable estimation of the qubit frequency variations by fitting deviations from a 2 MHz oscillation frequency. \textbf{b} All oscillations collected for qubit 6 in the linear device during a single collection run. \textbf{c} Deviations away from 2 MHz of the frequencies extracted from panel \textbf{b} as a function of time (left) and displayed as a histogram (right). The red line in the histogram is a fit to a Gaussian distribution, with a fitted standard deviation of $\sigma=33.9$ kHz. }
    \label{fig: experimental setup}
\end{figure}

\subsection*{Averaged 1-hour qubit frequency variations extracted from Ramsey-style experiments}

In order to quantify qubit frequency variations during the stable periods more accurately and on shorter timescales, we utilize Ramsey experiments that are implemented overnight to extract the average 1-hour qubit frequency variations experienced by all qubits in both the linear and 2$\times$2 devices. The circuit we implement and an example Ramsey oscillation are shown in Fig.~\ref{fig: experimental setup}a shows schematically the circuit used to perform a Ramsey experiment where a virtual $R_z(\omega t)$ operation, also referred to as virtual detuning, is included to facilitate the data fitting and analysis. A single Ramsey experiment is shown in the plot below, which takes approximately 30 seconds to run. We repeat these measurements for 10-41 hours resulting in a full dataset as shown in Fig.\ref{fig: experimental setup} b.

A total of 46 such datasets were collected, with 24 and 22 datasets collected from the linear and 2$\times$2 devices respectively (further details are provided in Supplementary Section~\ref{sup: data table}). Operating conditions such as the mixing chamber temperature, gate voltages and magnetic field were kept constant throughout a single data collection run, however the temperature and gate voltages can vary from day to day.

\begin{figure*}[ht]
    \centering
    \includegraphics[width=\textwidth]{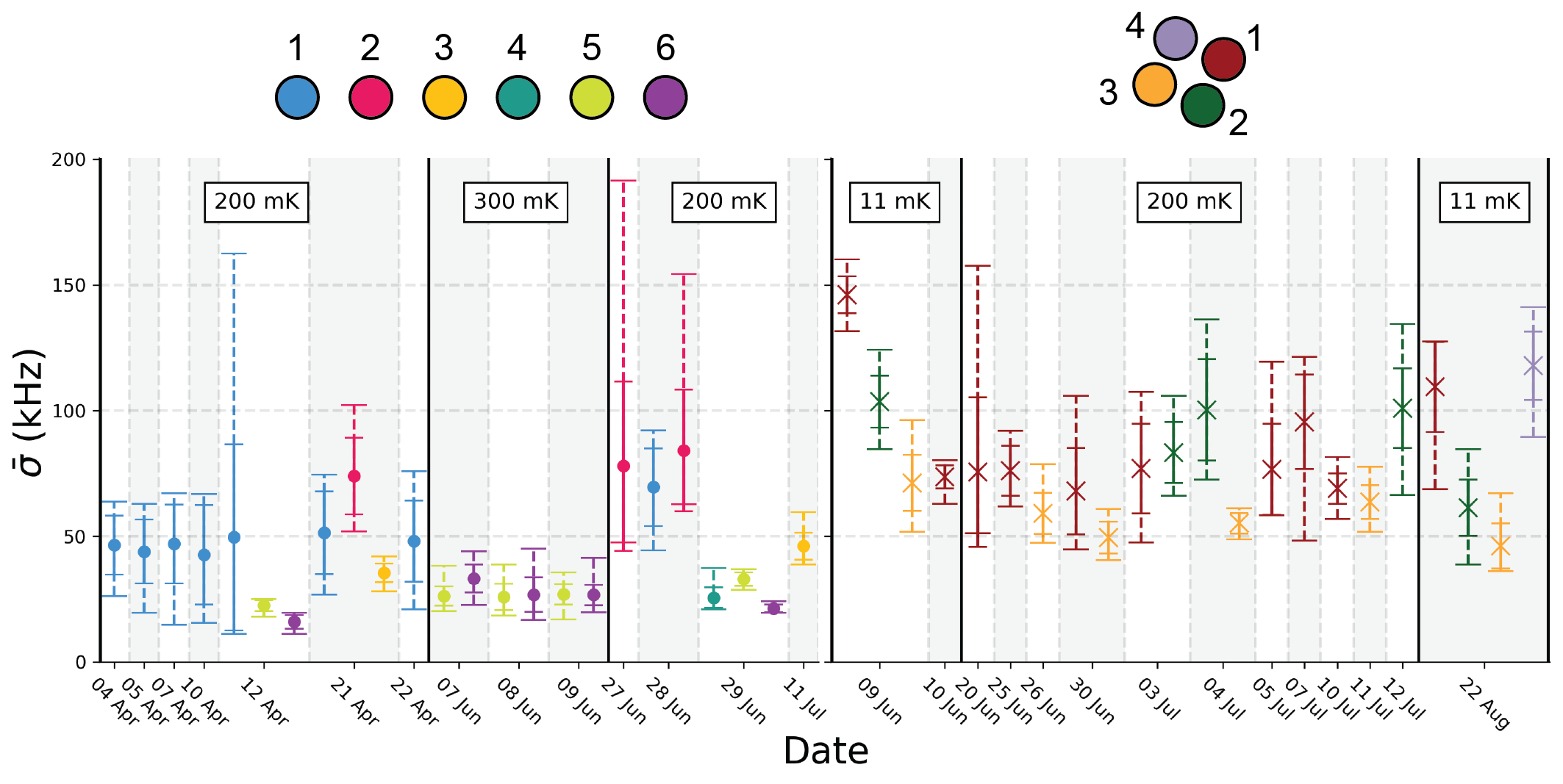}
    \caption{\textbf{Average 1-hour qubit frequency variation.} For each dataset, labeled by the day of the year in 2023, we segment the qubit frequency drift data into 1-hour sections and track the standard deviation of the recorded qubit frequencies $\sigma_i$ for each section $i$. The data points in the figure show the average of the standard deviation of the qubit frequency over one hour, $\bar{\sigma}$, for each dataset as described in Eq.~\ref{eq: avg sigma}. The solid error bars correspond to one standard deviation on $\sigma_i$ for each dataset (Eq.~\ref{eq: error sigma}) and express how the magnitude of the qubit frequency variations itself fluctuates throughout the data collection period (of 10-41 hours). The dashed error bars correspond to the minimum and maximum observed $\sigma_i$ throughout the data collection period (Eq.~\ref{eq: min max sigma}). The mixing chamber temperature at the time of data collection is displayed at the top of each panel.}
    \label{fig: freq drift deviation}
\end{figure*}

For most Ramsey experiments, the detuning was set to 2 MHz and the maximum wait time of 1.5$\mathrm{\mu}$s ensured a high degree of accuracy in fitting the oscillation frequency (for a few datasets, a 4 MHz detuning and 0.75$\mathrm{\mu}$s wait time were used). In the linear device, the average recorded uncertainty in the extracted qubit frequency is 8.7 kHz for the Ramsey experiments, compared to an average recorded uncertainty in the fit from the qubit frequency calibration scan of 22.0 kHz. These values are extracted from the covariance matrix of the fit to Eqs.~\ref{eq: Rabi formula} and~\ref{eq: Ramsey} (below) respectively and are averaged across all recorded measurements over all qubits. The number of data points in the Ramsey measurements was chosen to avoid aliasing. Readout calibrations were periodically interleaved every 10 Ramsey experiments to ensure high visibility throughout the full collection period. When data was collected on multiple qubits, a full Ramsey trace was obtained on one qubit before proceeding to the next. Finally, each oscillation is fitted with a function of the form

\begin{equation}\label{eq: Ramsey}
    p(t) = \frac{A}{2}\text{cos}(2\pi\tilde{\Delta} t + \phi)e^{-(t/T_2^*)^2} + B \;,
\end{equation}
where the fit parameters are the oscillation frequency $\tilde{\Delta}$, a phase offset $\phi$, the dephasing time $T_2^*$, and parameters $A$ and $B$ that depend on state preparation and readout errors. Deviations of $\tilde{\Delta}$ away from 2 MHz correspond to deviations in the qubit frequency.  We assume all of the used fitting parameters to be quasi-static during a single Ramsey experiment, justifying the use of a Gaussian decay envelope. After fitting the oscillations, we perform a filtering protocol utilizing a normalized chi-square metric to filter out any low-quality traces (as described in Supplementary Section~\ref{sup: filtering}) to form a dataset as shown in Fig.~\ref{fig: experimental setup}c.

In Fig.~\ref{fig: freq drift deviation}, we plot the average of the qubit frequency variations $\hat{\sigma}$ observed over 1-hour windows. Each full qubit frequency drift dataset, consisting in total of $N$ datapoints $\mathbf{x}=(x_1, x_2, ... , x_N)$, is segmented into $M$ chronological sections $[\mathbf{x}^1,\mathbf{x}^2,...,\mathbf{x}^M]$ where $M$ was selected such that each $\mathbf{x}^i$ would cover approximately a 1-hour time window. For every section $i$ the corresponding standard deviation $\sigma_i$ in qubit frequency was calculated. The datapoints plotted in Fig.~\ref{fig: freq drift deviation} show the average of those standard deviations for a given dataset $\mathbf{x}$, i.e:
\begin{equation}\label{eq: avg sigma}
    \bar{\sigma} = \frac{1}{M}\sum_{i=1}^M\sigma_i \;.
\end{equation}
From the frequency drift values $\sigma_i$ of the $M$ one-hour sections, we also extract their standard deviation, and the range from the minimum to the maximum value, i.e: 
\begin{equation}\label{eq: error sigma}
    \sigma_{\text{std}} = \sqrt{\frac{\sum_i(\sigma_i - \bar{\sigma})^2}{M}},
\end{equation}
\begin{equation}\label{eq: min max sigma}
    \sigma_{\text{min}}=\min\limits_i\sigma_i\;,\;
    \sigma_{\text{max}}=\max\limits_i\sigma_i \;,
\end{equation}
which are plotted as error bars in Fig.~\ref{fig: freq drift deviation}. We do not have enough data points to draw conclusions on the impact of the mixing chamber temperature on the average 1-hour qubit frequency drifts.

For all measurements, the qubit frequencies show a maximum standard deviation of less than 200 kHz within a 1-hour time window. Focusing first on the linear device, we observe that $\sigma_\text{max}$ of qubits 4-6 never exceeds 50 kHz within a single data collection period. However, we can also notice that qubits 1 and 2 exhibit both a larger average and spread in $\sigma_i$ compared to the rest, highlighting how the frequency variations in time of the qubits can vary substantially within the same device.

We speculate that the larger frequency variations seen in qubits 1 and 2 are due to localized and strongly-coupled TLFs, which could originate from electric~\cite{Elsayed22} (e.g. from nearby charge defects in the heterostructure or gate dielectrics) or magnetic~\cite{Hensen2020} (Si$^{29}$ nuclear spins) sources. For example, we can identify the presence of a TLF strongly coupling to qubit 1 which induces a telegraph signal in the qubit frequency with an amplitude of 103-315 kHz and a switching rate on the order of twice per hour (Supplementary Fig.~\ref{sup fig: TLF behaviour}). We note that the activation and switching rate of individual charge TLFs are known to be affected by DC voltages~\cite{Ye24A,Ye24B}. Given that the separate datasets $\mathbf{x}$ are typically taken with different DC gate voltage settings, this can explain the variability in the observed $\bar{\sigma}$, $\sigma_\text{min}$ and $\sigma_\text{max}$ for any given qubit. Furthermore, in the course of collecting full datasets $\mathbf{x}$ lasting between 10 and 41 hours, we have also observed three instances of a single discrete jump in qubit frequencies of about 250 to 380 kHz in amplitude for qubits 1 and 2, which explain the large $\sigma_\text{max}$ on 12/04, 27/06 and 28/06. These are documented further in Supplementary Section ~\ref{sup: TLF and discrete jumps}. 

In the 2$\times$2 device, we observe that qubit 3 consistently experiences lower levels of frequency variations when compared with qubits 1 and 2. Furthermore, we observe larger deviations in all qubit frequencies when compared with the linear device. A priori, we could expect qubits from both devices to exhibit similar levels of drift, since both devices follow nearly identical fabrication procedures and are hosted in similar heterostructures (detailed in the Methods section). In fact, the 2x2 device has a 1nm thinner quantum well which has been shown to reduce the strength of charge noise~\cite{Paquelet23}. 

A major difference between the two devices however is the micromagnet gradient field at the quantum dot locations. In the linear device, the micromagnet was engineered such that the qubits would be positioned in the sweetspot of the decoherence gradient. When moving to two dimensions in the 2$\times$2 device, the qubits no longer reside in the sweetspot. Whereas the exact dot locations in the actual device are not known, we expect the resulting decoherence gradients to be roughly twice larger than in the linear device, based on numerical simulations~\cite{Philips22, Unseld23}. This could induce larger qubit frequency variations in the 2$\times$2 device even if the level of charge noise were the same or lower as in the linear device.

\subsection*{Fidelity of $X_{\pi/2}$ gate impacted by qubit frequency detuning}

\begin{figure}[ht]
    \centering
    \includegraphics[width=\linewidth]{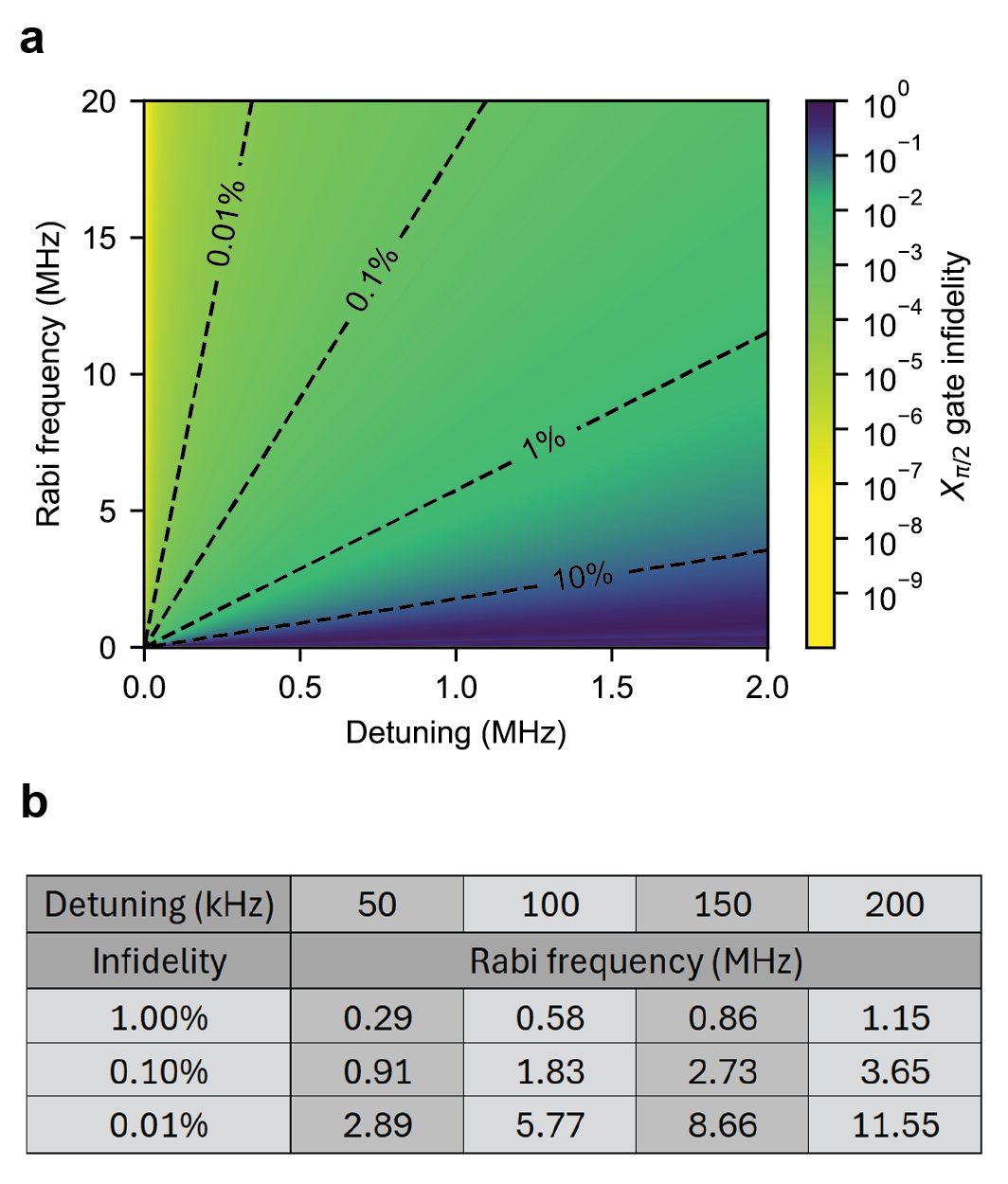}
    \caption{$\mathbf{X_{\pi/2}}$ \textbf{gate infidelity}. \textbf{a} The infidelity of an $X_{\pi/2}$ gate, in the absence of other sources of noise, as a function of qubit frequency detuning and Rabi frequency. Dashed black lines represent the gate infidelity corresponding to 0.01\%, 0.1\%, 1\% and 10\%. \textbf{b} Minimum Rabi frequency needed to retain 1\%, 0.1\% and 0.01\% infidelity at the specified qubit frequency detunings.} 
    \label{fig: detuning vs fidelity}
\end{figure}

The uncertainty in the qubit frequencies at any point in time has several consequences. First, when a qubit is idling in a state of superposition, random fluctuations in the resonance frequency will result in a random phase pickup. The effect of slow qubit frequency variations during idling can be largely removed using dynamical decoupling techniques based on the spin-echo concept, although the decoupling pulses themselves will also produce errors~\cite{Burkard23}. Second, when attempting to resonantly drive a qubit, the drive frequency will no longer be on resonance if the qubit frequency has shifted. In the literature, this effect has been mitigated by background recalibrations of the qubit frequency and adjusting the drive frequency accordingly~\cite{Philips22}. More resilience to unknown qubit frequency detunings can be achieved using composite rotations or by tailoring the amplitude and/or phase profile of the drive signal~\cite{Vandersypen04}. Nevertheless, it is instructive to quantify how qubit frequency detunings affect the fidelity of single-qubit gates implemented by standard fixed-amplitude and fixed-frequency bursts.

The EDSR Hamiltonian (Eq.~\ref{eq: EDSR rotating frame}) enables us to engineer arbitrary rotations about an axis in the \textit{xy}-plane of the Bloch sphere by adjusting the amplitude, phase and duration of the microwave burst. We here assume that the burst is applied nominally on resonance. However, the presence of qubit frequency noise $\eta(t)$, which we take to be quasi-static for the duration of a burst, introduces a detuning $\Delta(t)=\eta(t)$, i.e. a random $\hat{\sigma}_z$ component which effectively reduces the control fidelity if left unaccounted for.  We can derive the expected gate fidelity~\cite{pedersen2007fidelity} of a single-qubit $X_{\pi/2}$ gate (Supplementary 
Section~\ref{sup: gate fidelity}) as a function of the detuning $\Delta$ and the Rabi frequency $\Omega$ as 
\begin{equation}\label{eq: Xpi/2 fidelity}
    F(\Delta,\Omega) = \frac{1}{3}\Bigl[1 + \frac{\Omega^2}{\lambda^2} + \frac{\Delta^2}{\lambda^2}\text{cos}^2\Bigl(\frac{\lambda\pi}{4\Omega}\Bigr)+\frac{\Omega}{\lambda}\text{sin}\Bigl(\frac{\lambda\pi}{2\Omega}\Bigr)\Bigr] \;,
\end{equation}
with $\lambda=\sqrt{\Delta^2+\Omega^2}$.

In Fig.~\ref{fig: detuning vs fidelity}a we plot the $X_{\pi/2}$ gate infidelity as a function of the Rabi frequency and qubit frequency detuning. As is well known, for a given frequency detuning, we can enhance the fidelity by increasing the Rabi frequency through a larger driving amplitude. In Fig.~\ref{fig: detuning vs fidelity}b, we show a table with the required Rabi frequencies corresponding to an $X_{\pi/2}$ gate infidelity of 1\%, 0.1\% and 0.01\% at set qubit frequency detunings of 50, 100, 150 and 200 kHz. In the linear device, the frequencies of qubits 5 and 6 do not vary by more than 50 kHz in most 1-hour time windows. According to Eq.~\ref{eq: Xpi/2 fidelity}, at this level of detuning and to retain $X_{\pi/2}$ gate fidelities of 99.99\%, it is sufficient to drive with a Rabi frequency of 2.89 MHz. The frequency of qubit 3 in the 2x2 device varies by less than 100 kHz for most one-hour windows, and thus it is sufficient to drive at 5.77 MHz to retain the same level of fidelity. However, long-term drift on the scale as observed in Fig.~\ref{fig: stable periods} clearly cannot be left uncompensated as, for example, a detuning of 0.5 MHz would require a Rabi frequency of 28.9 MHz to retain a $X_{\pi/2}$ gate fidelity of 99.99\%, which is unattainable in this device. As also observed in~\cite{Yoneda2018}, the Rabi frequency becomes non-linear with respect to the microwave amplitude and eventually saturates. Furthermore, randomized benchmarking experiments have indicated that the highest attainable Rabi frequencies do not necessarily produce the highest fidelities~\cite{Noiri22}, since heating and leakage effects become more severe at larger driving amplitudes.

\begin{figure*}[ht]
    \centering
    \includegraphics[width=\textwidth]{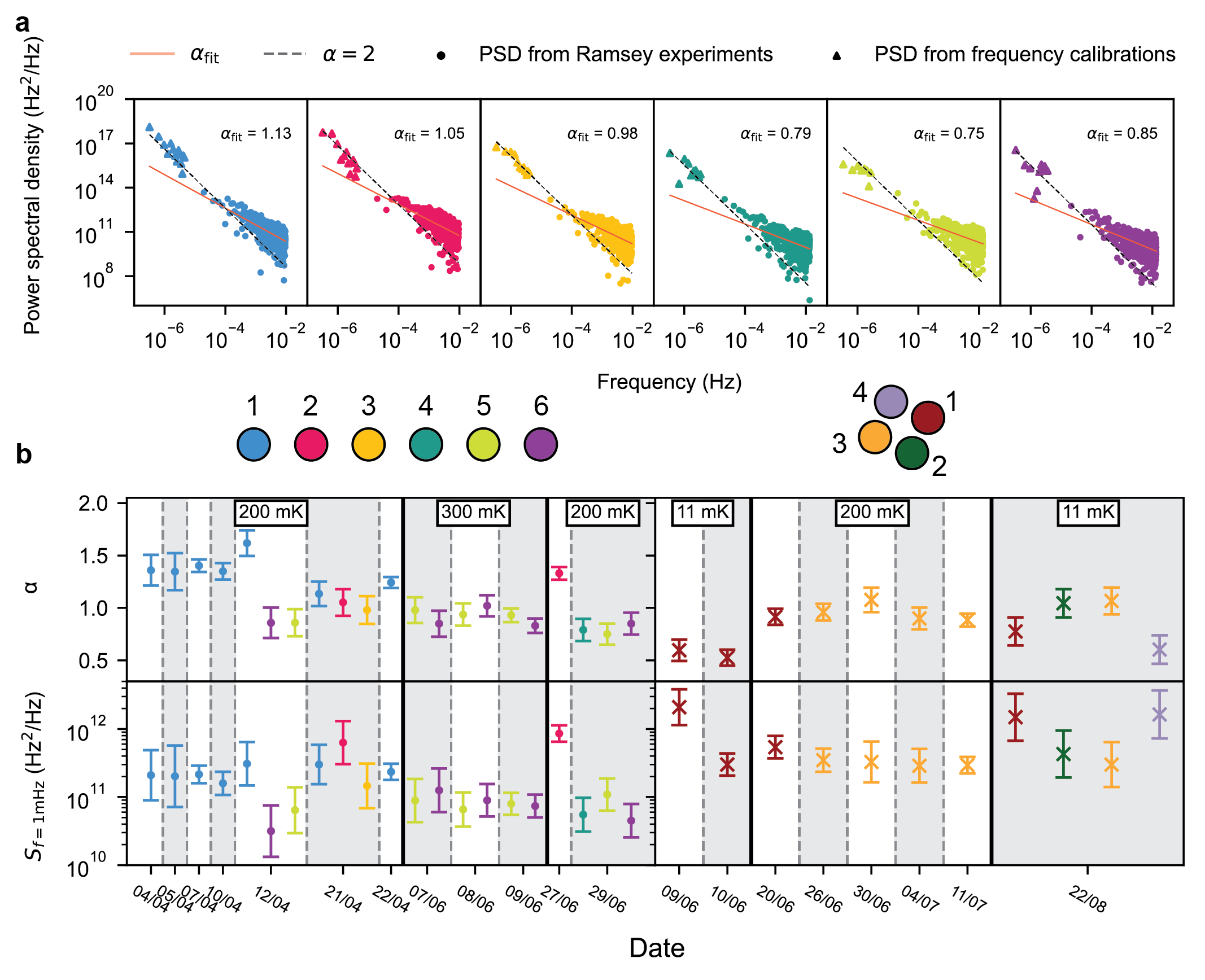}
    \caption{\textbf{Characterization of low-frequency qubit frequency noise spectra}. \textbf{a} A comparison of the PSDs extracted from the Ramsey data and the frequency scan data for qubits 1-6 in the linear device. The frequency calibration data is from period 10 as described in Fig.~\ref{fig: stable periods}, which corresponds to the longest time period with a sufficient number of datapoints at a stable operating condition of gate voltage, magnetic field and temperature. The Ramsey data for qubits 1-3 was collected on 21/04/2023 and for qubits 4-6 on 29/06/2023. The solid red lines are fits to the Ramsey PSD data points to $S_0/f^\alpha$, with $S_0$ and $\alpha$ as fit parameters. The black dashed line is a guide to the eye showing a $1/f^2$ dependence. We note an unexpected change in slope of the PSD from $1/f^\alpha$ (with $\alpha \approx1$) to $1/f^2$ at a transition frequency of approximately $10^{-4}$ Hz for all qubits. \textbf{b} Fitted values of $\alpha$ and $S_\text{f=1mHz}$ of each PSD obtained from fits to the Ramsey data as in \textbf{a}. The DC gate voltage setting for each date is reported in Supplementary Section \ref{sup: DC logs}. The error bars (2$\sigma$) are extracted from the covariance matrix of the fit. The mixing chamber temperature at the time of data collection is shown at the top.} 
    \label{fig: PSDs}
\end{figure*}

\subsection*{Qubit frequency noise spectra}

In order to get more insight in the nature of the qubit frequency variations, we perform Fourier analysis on the data collected in the time domain to obtain the noise power spectral density (PSD). The Ramsey-style data are used to extract the noise spectra over a frequency range of approximately $[10^{-4},10^{-2}]$ Hz, whilst the frequency scan  data are used to probe the noise spectra over a frequency range of approximately $[10^{-7},10^{-5}]$ Hz. Even lower frequency ranges were not accessible despite the total data collection period of 912 days, as we only consider data taken at the during the specified stable periods from Fig.~\ref{fig: frequency log}. This gave us datasets up to a timescale of 36 days, yielding noise spectra in the sub-$\mathrm{\mu}$Hz range. Even studies on the timescale of a month are rare, so experimental data available in the literature for comparison remains limited.

Qubit frequency noise with a $1/f^\alpha$ spectrum, with $\alpha\approx1$, has been reported in Si/SiGe devices~\cite{Yoneda2018, Struck20, Yoneda23, Rojas23, Rojas24}. However, these studies characterize the noise spectra only at frequencies of $10^{-4}$ Hz and above, with the exception of~\cite{Struck20}. Interestingly, that study reports a cross-over in the qubit frequency noise spectrum from $\alpha \approx 2$ below $f\approx 10^{-4}-10^{-3}$ Hz, to $\alpha \approx 1.5$ at higher frequencies, and notes a corresponding transition in the charge noise spectrum from $1/f^2$ to $1/f$. A transition in the charge noise spectrum with similar characteristics is reported in~\cite{Elsayed22} for Si/SiO$_2$ quantum dots as well.

In Fig.~\ref{fig: PSDs}a, we plot the PSD data points from both the Ramsey experiments (circles) and the frequency scans (triangles) (further details on the methods used to obtain these plots are provided in the Methods and Supplementary Section~\ref{sup: PSDs}). The solid red lines are fits to the PSD data points from the Ramsey measurements. The fits give $\alpha$ in the range of $0.68 - 1.62$. We can observe that the PSDs extracted from frequency calibration data consistently lie above the trends extrapolated back from the PSDs extracted from Ramsey data, indicating a transition to a PSD with a larger value of $\alpha$ at lower frequencies. The black dashed lines are a guide to the eye corresponding to $\alpha=2$, and visually match the observations reasonably well. The transition frequency is around $f=10^{-4}$ Hz, not very different from the earlier observations~\cite{Struck20,Elsayed22}. In principle, $1/f$ noise spectral densities could result from the evolution of a sparse bath of nuclear spins. However, the fact that also charge noise transitions to $1/f^2$ behavior below $f=10^{-4}$ Hz~\cite{Struck20,Elsayed22}, hints that the qubit frequency noise at these frequency ranges could arise from atypical charge noise, i.e. not originating from the ensemble of TLFs with log-uniformly distributed switching rates that gives rise to the typical $1/f$ behavior. 

In Fig.~\ref{fig: PSDs}b, we show the fitted parameters of the PSDs from all Ramsey experiments to $S(f)=S_0/f^\alpha$. Most qubit frequency noise spectra exhibit a good agreement to the fit, although the fitted values of $\alpha$ show a large degree of variability. We observe consistently lower values of $S_\text{f=1mHz}$ in the linear device when compared to the 2x2 device, which was anticipated given the smaller qubit frequency variations seen in the time domain (see Fig.~ \ref{fig: freq drift deviation}). The observed values for $S_\text{f=1mHz}$ are in the same range as the values reported in earlier studies~\cite{Yoneda2018, Struck20, Yoneda23, Rojas23, Rojas24}, which range from approximately $[10^{11},10^{13}]$ Hz$^2$/Hz.

%with the exception of qubit 1 and 2 from the linear device exhibiting an added Lorentzian component $\propto 1/f^2$ due to a strong coupling to a TLF, which in our analysis leads to an unreliable estimation of the $\alpha$ parameter and thus have been omitted from the presented results, however these figures are presented in Supplementary Section~\ref{sup-sup: PSDs}. The results presented in the main text encompass data from 8 different qubits from both the linear and 2x2 devices.

Most noise profiles show relative consistency over time, but at the same time the variability seen in Fig.~\ref{fig: PSDs}b highlights that caution is due when drawing firm conclusions on the value of $\alpha$ or $S_\text{f=1mHz}$ taken on any particular day. A strong correlation between qubit frequency noise and charge noise was documented in~\cite{Yoneda23}. Furthermore, operating conditions such as temperature~\cite{Connors19} can influence charge noise levels and DC gate voltages~\cite{Ye24A} can (de)activate individual TLFs, and it is likely that both will also influence qubit frequency noise spectra. This being said, we also observe changes in the extracted values of  $\alpha$ or $S_\text{f=1mHz}$ under identical operating conditions of temperature, DC voltages and magnetic field. This is the case for qubit 1 in the 2x2 device from June 9th - 10th and to a lesser extent for qubit 6 of the linear device from June 7th-9th. The same behavior is seen in the time-domain data in Fig.~\ref{fig: freq drift deviation}.

\section*{Discussion}
We have quantified the qubit frequency variations and characterised the PSDs of 10 different qubits across two different devices. In some cases, such as qubits 1 and 2 of the linear device, the frequency variations exhibit telegraph signals and are presumably dominated by a strong coupling to a local TLF. Such events are particularly harmful for maintaining high-fidelity qubit operations. In the PSDs, the most striking observation is a transition in the slope of the qubit frequency noise spectrum for all qubits in the linear device from $1/f$-like behavior above $\approx 10^{-4}$ Hz to $1/f^2$-like behavior at frequencies below $\approx 10^{-4}$ Hz.

Whereas it is difficult to be certain of the origin of the TLF's and the observed power spectral densities, we can consider the following. For the sparse bath of nuclear spins present in the device (under 800 ppm residual $^{29}$Si in the quantum well), the magnetic noise spectrum can tend towards a $1/f^\alpha$-like slope with $\alpha < 2$~\cite{Eng15,Madzik20,Burkard23}. Furthermore, typical nuclear spin bath models assume that the low-frequency tail of the spectrum flattens out~\cite{Rojas24}. Considering this, it is unlikely that the transition that we observe is caused solely by the nuclear spin bath. As for two-level charge fluctuators, any noise spectrum can in principle be explained by a certain distribution of the switching rates of the TLFs. For example, a $1/f^2$ spectrum above a certain frequency and a flat spectrum below it, points at a single dominant TLF, whereas a $1/f$ spectrum is indicative of a log-uniform distribution of switching rates. For the same reason, a combination of nuclear spin noise and electric TLFs cannot be excluded given the observed qubit frequency PSDs. However, it is worth noting that a similar transition from $1/f^2$ to $1/f$ behavior, and with a comparable cross-over frequency, has been observed in charge noise spectra reported in \cite{Elsayed22} and \cite{Struck20}. The similarity would suggest an electrical origin of the observed transition in the qubit frequency PSDs, which may be caused by a distribution of charge TLFs that deviate from the more typical log-uniformly distributed switching rates. Another possible explanation could be the presence of mobile charges in the gate oxide~\cite{Elsayed22}. These charges are free to slowly move around and could be expected to exhibit Brownian motion which has a characteristic noise spectrum with a $1/f^2$ slope. In this case, the picture could be that Brownian motion of slowly moving charges dominates the PSD below $\approx 10{-4}$ Hz and a standard log-uniform distribution of TLFs dominates the PSD at higher frequencies, with additional contributions from nuclear spin noise.

Whereas we cannot be sure of the dominant sources of frequency variations, we can still recommend several steps to obtain more stable qubit frequencies. Nuclear spin noise can be further suppressed by eliminating additional $^{29}$Si nuclear spins in the quantum well (recent studies have already achieved 50ppm residual $^{29}$Si in the quantum well~\cite{Klos24, Huang24}), and by also isotopically purifying the SiGe barrier into which the electron wavefunction penetrates, reducing both the number of $^{29}$Si and $^{73}$Ge nuclear spins. For charge fluctuators, it is commonly believed that the majority of charge traps reside in the gate oxide layer. Reducing the thickness of this layer has shown to be effective in reducing charge noise levels at the quantum well \cite{Connors19}. A systematic effort to develop gate oxides that minimize charge noise at cryogenic temperatures will be needed to obtain a further reduction in the qubit frequency variations. Furthermore, the design of the micromagnet can also be optimized further to decrease the decoherence gradient, which will directly reduce the impact of electrical TLFs on the qubit frequency, albeit with a trade-off in qubit addressability and possibly the driving efficiency. Finally, we note that reducing the possible mechanisms for qubit frequency variations one at a time, can help clarify what are the dominant mechanisms.

\section*{Conclusion}
We report the calibrated qubit frequencies for all six qubits in a linear quantum dot device over a span of 912 days. Over this entire period, the qubit frequencies varied by up to $\pm 100$ MHz, under the influence of changes in the DC gate voltages and mixing chamber temperature, and while running a variety of experiments on the device. We identified 16 time windows of one week or longer during which the gate voltages were not adjusted by no more than 15 $\mathrm{\mu}$V, yet qubit initialization, control and readout were maintained. During those ``stable'' periods the qubit frequencies varied by less than $\pm  7$ MHz. On days where the gate voltages, magnetic field and temperature were all kept constant, we performed overnight Ramsey measurements on the linear device and a 2x2 device. Over 1-hour time windows during these overnight measurements, the standard deviation in the qubit frequency was below 200 kHz for all qubits, and below 50 kHz on some qubits of the linear device. To put this number in perspective, a detuning of the qubit frequency by 50 kHz permits a 99.99$\%$ single-qubit gate fidelity provided the Rabi frequency exceeds 2.9 MHz, which is an accessible value in practice. We also converted the time-domain data to power spectral densities for the qubit frequencies, fitting and comparing with a $S(f)=S_0/f^\alpha$ dependence. Since the physically possible frequencies are bounded, one might expect the PSD to saturate at very low frequencies. Instead, we observe that the PSD, hence the amplitude of the qubit frequency variations, continues to increase all the way down to the $\mathrm{\mu}$Hz range (corresponding to weeks). Most strikingly, comparing the noise spectra from the Ramsey and frequency calibration data revealed in fact a steeper slope $\alpha$ at frequencies below $\approx 10^{-4}$ Hz.  Furthermore, we observed that the noise PSDs not only vary substantially between two devices despite the similar heterostructure growth conditions and fabrication flow used, but also between qubits within the same device, and in some cases show variability from day to day. 

Whereas dedicated studies of the stability of qubit parameters on the timescale of weeks to months obviously require patience to carry out, it will be important to continue such experiments as it looks problematic to extrapolate qubit frequency variations on this ultra-long timescale from measurements on more accessible timescales. The better the long-term stability, the lower the demands on background calibrations will be, and the simpler the autonomous operation of spin-qubit systems. It will be especially helpful to eliminate dominant two-level fluctuators, which cause the qubit frequency to jump by a large amount (of the order of 103-351 kHz) at rare but random points in time.

\section*{Methods}

\subsection*{Device heterostructure and fabrication}

Both devices are fabricated on a $^{28}$Si/SiGe heterostructure, with a strain-relaxed $\text{Si}_{0.7}\text{Ge}_{0.3}$ buffer layer. On top of this an isotopically enriched $^{28}$Si quantum well is grown (residual $^{29}$Si concentration below 800 ppm) with a thickness of 8~nm and 7~nm for the 6D2S and 2x2 devices respectively. Then a 30~nm thick Si$_{0.7}$Ge$_{0.3}$ buffer layer is grown with a 1~nm silicon oxide cap. A gate stack consisting of three layers of Ti:Pd metallic gates is deposited and separated from the heterostructure by Al$_2$O$_3$, and separated from each other by 5~nm Al$_2$O$_3$. The metal gate film thickness is as described in~\cite{Lawrie20}. Finally, another Al$_2$O$_3$ layer is deposited and a 200~nm thick Ti:Co micromagnet is fabricated.

\subsection*{Obtaining power spectral densities}

We note that the extraction of PSDs from the time domain data is hindered by both a low number of datapoints and a non-uniform sampling rate. To overcome this, we first interpolate our time-domain data onto a uniform time axis before implementing the standard FFT. We validated this procedure by performing an analysis on artificially generated data (see Supplementary Section~\ref{sup: freq calib PSD} for details).

\section*{Acknowledgements}

We acknowledge the contribution of A. Sammak in the growth of the $^{28}$Si/SiGe heterostructures. We also thank J. Rojas-Arias, P. Stano and the members of the Vandersypen group for useful discussions. This research was supported by the Horizon Europe program of the European Union under grant agreement no. 951852 (QLSI) and no. 101174557 (QLSI2), by the “Quantum Inspire–the Dutch Quantum Computer in the Cloud” project (Project No. NWA.1292.19.194) of the NWA research program “Research on Routes by Consortia (ORC),” which is funded by the Dutch Research Council (NWO), by Intel Corporation and by the Army Research Office (ARO) under grant numbers W911NF-17-1-0274 and W911NF-12-1-0607. The views and conclusions contained in this document are those of the authors and should not be interpreted as representing the official policies, either expressed or implied, of the ARO or the US Government. The US Government is authorized to reproduce and distribute reprints for government purposes notwithstanding any copyright notation herein. 

\section*{Data availability}

The data and code used throughout this manuscript are available for access in the Zenodo repository (https://doi.org/10.5281/zenodo.15632269). 


% \begingroup
%   \renewcommand{\addcontentsline}[3]{}% temporarily disable ToC writes
%   \bibliographystyle{apsrev4-2}      % or your preferred style
%   \bibliography{refs_utf8}                 % your .bib file (no .bib extension)
% \endgroup

\clearpage

\twocolumngrid
\onecolumngrid
\appendix 

\clearpage
\onecolumngrid         % make sure we are in one-column mode
\thispagestyle{empty}  % suppress headers/footers on the title page

\begin{center}
  \vspace*{2cm}        % push down from the top margin

  {\Large\bfseries Supplementary Information for tracking spin-qubit frequency variations over 912 days\\[6pt]}

  Kenji Capannelli$^1$,
  Brennan Undseth$^1$,
  Irene Fernández de Fuentes$^1$,
  Eline Raymenants$^1$,
  Florian K. Unseld$^1$,\\
  Oriol Pietx-Casas$^1$,
  Stephan G. J. Philips$^1$,
  Mateusz T. Mądzik$^1$,
  Sergey V. Amitonov$^2$,\\
  Larysa Tryputen$^2$,
  Giordano Scappucci$^1$,
  Lieven M. K. Vandersypen$^1$\\[8pt]

  {\itshape 1 QuTech and Kavli Institute of Nanoscience, Delft University of Technology,\\
  Lorentzweg 1, 2628 CJ Delft, The Netherlands}\\[2pt]
  {\itshape 2 QuTech and TNO, Stieltjesweg 1, 2628 CK Delft, The Netherlands}\\[16pt]

\end{center} 

\renewcommand{\thesection}{S\arabic{section}}
\renewcommand{\thefigure}{S\arabic{figure}}
\renewcommand{\theequation}{S\arabic{equation}}
\setcounter{equation}{0}
\setcounter{section}{0}
\setcounter{figure}{0}

\title{Supplementary information for tracking spin qubit frequency variations over 912 days}
\author{Kenji Capannelli}
\author{Brennan Undseth}
\author{Irene Fern\'{a}ndez de Fuentes}
\author{Eline Raymenants}
\author{Florian~K.~Unseld}
\author{Oriol Pietx-Casas}
\author{Stephan~G.~J.~Philips}
\author{Mateusz~T.~M\k{a}dzik}
\affiliation{QuTech and Kavli Institute of Nanoscience, Delft University of Technology, Lorentzweg 1, 2628 CJ Delft, The Netherlands}
\author{Sergey~V.~Amitonov}
\author{Larysa Tryputen}
\affiliation{QuTech and Netherlands Organization for Applied Scientific Research (TNO), Stieltjesweg 1, 2628 CK Delft, The Netherlands}
\author{Giordano Scappucci}
\author{Lieven~M.~K.~Vandersypen}
\affiliation{QuTech and Kavli Institute of Nanoscience, Delft University of Technology, Lorentzweg 1, 2628 CJ Delft, The Netherlands}
 
\maketitle

\tableofcontents

\newpage 

\section{Data fitting and filtering}\label{sup: filtering}
\subsection{Normalised chi-square}
To evaluate the quality of the fits to both the qubit frequency calibration and Ramsey-like experiment datasets, we utilise a normalised chi-square metric. For a dataset $\mathbf{p}=[p_1,...,p_n]$, we fit it to its corresponding model $\mathbf{P}=[P_1,...,P_n]$, which are given by equations~\ref{eq: Rabi formula} and~\ref{eq: Ramsey} in the main text for the frequency calibration and Ramsey datasets respectively. The chi-square metric is then given by

\begin{equation}
    \chi^2 = \frac{1}{n}\sum_{i=1}^n(P_i-p_i)^2
\end{equation}

However, when running thousands of experiments, drifts in readout and initialization parameters will cause fluctuations in the visibility of the experimental data. To account for this, the visibility is also included as a fitting parameter. However, this can at times result in a bad quality fit with a high overall chi-square score for otherwise valid experimental data. The converse is also true, in that an erroneous scan might return an unexpectedly low chi-square score. To ensure that our filtering process overcomes these extreme cases, we adapt our metric to the \textit{normalised} chi-square metric. Normalising the data ensures a fair evaluation of the quality of the fit for oscillations with suppressed visibility. This is defined as the following

\begin{equation}
    \tilde{\chi}^2 = \frac{1}{n}\sum_{i=1}^n(\tilde{P}_i-\tilde{p}_i)^2
\end{equation}

where 
\begin{equation}
    \tilde{P}_i = \frac{P_i - p_\text{min}}{p_\text{max}-p_\text{min}}, \;\;\;\tilde{p}_i = \frac{p_i - p_\text{min}}{p_\text{max}-p_\text{min}}
\end{equation}
with 
\begin{equation}
    p_{\text{max}} = \max_ip_i, \;\;\;p_{\text{min}} = \min_ip_i
\end{equation}

We utilize this metric to evaluate the quality of fit for every single frequency calibration and Ramsey-like experiment. Any low-quality experimental data that achieves a normalized chi-square metric score of a specified threshold or higher is filtered out and the remaining data is retained for analysis.  
\clearpage 

\subsection{Additional filtering for frequency calibration data}

After evaluating the normalized chi-square score for every frequency calibration, all scans with a score of 0.2 or higher are filtered out. After this we perform an extra round of filtering to remove any datasets taken at different magnetic field settings. The datasets before and after this filtering step is illustrated below in Figs.~\ref{fig: before filtering} and~\ref{fig: after filtering}.

\begin{figure}[ht]
    \centering
    \includegraphics[width=0.85\linewidth]{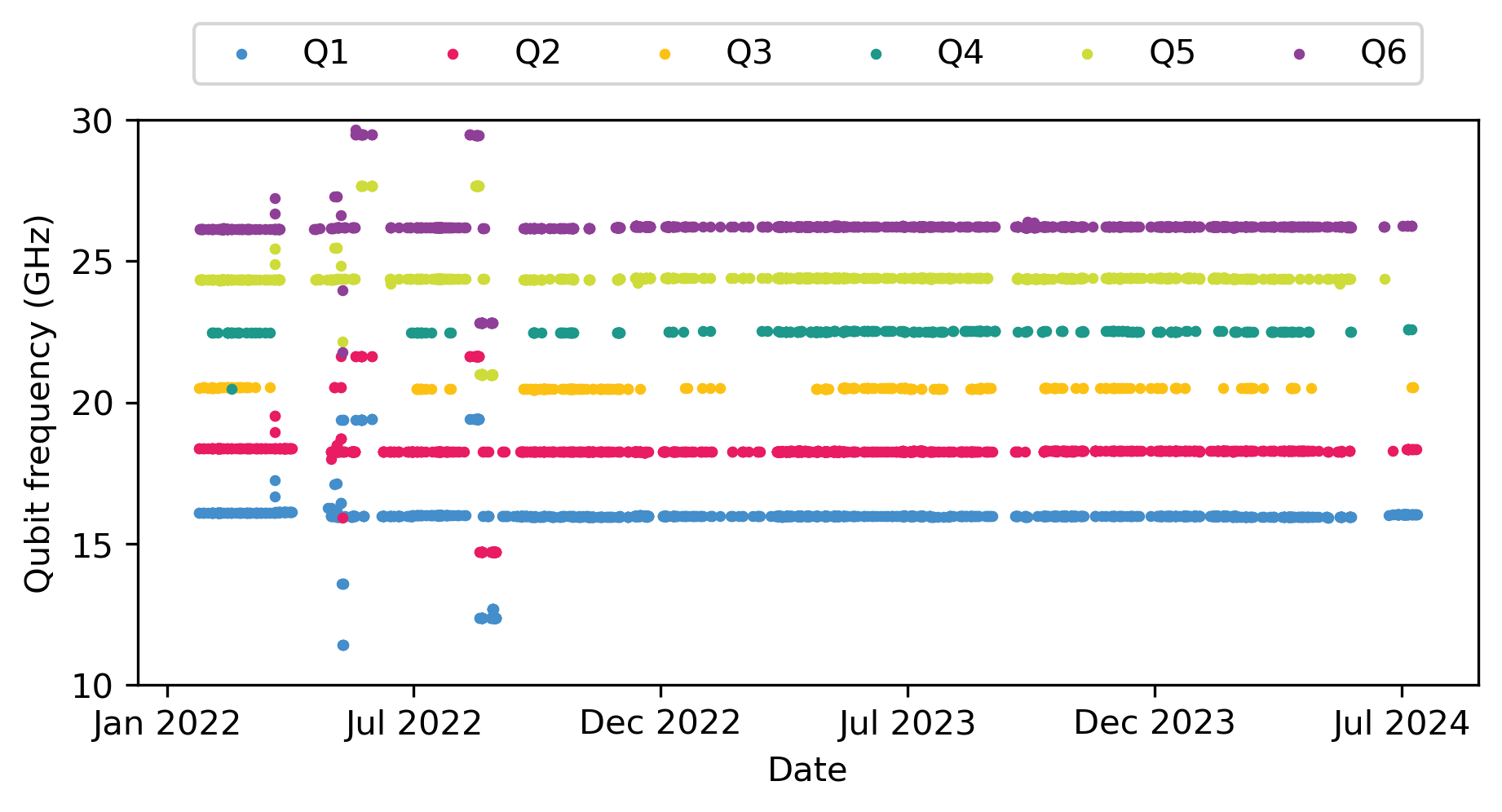}
    \caption{\textbf{Before additional filtering.} Qubit frequencies extracted from frequency calibrations scans over the time period of 24/01/22 - 24/07/24 after the filtering procedure utilizing the normalized chi-square metric as described above for all qubits in the linear device. Each trace is offset from each other by 2 GHz for visual clarity.}
    \label{fig: before filtering}
\end{figure}

\begin{figure}[ht]
    \centering
    \includegraphics[width=0.85\linewidth]{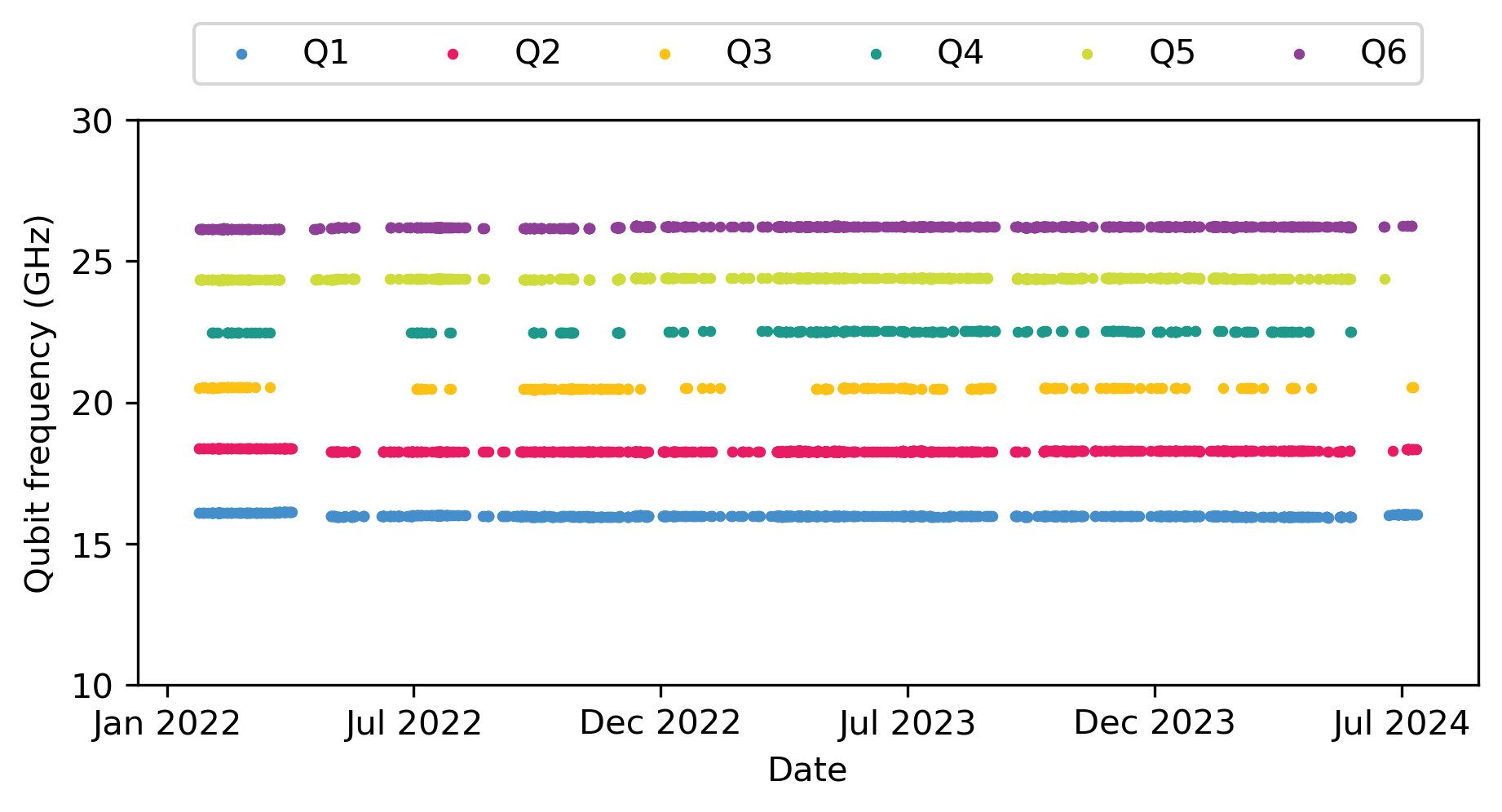}
    \caption{\textbf{After additional filtering.} After the filtering done with the normalized chi-square metric, we do an additional round of filtering to filter out any scans performed at magnetic field settings that were different to 400 mT.}
    \label{fig: after filtering}
\end{figure}

\clearpage 

\section*{Qubit frequency and gate voltage correlations from frequency calibration experiments}\label{sup: q freq vs gate voltage}

In this section we plot qubit frequency versus the DC voltage of each gate in the array for qubits 1, 2, 5 and 6, for the blue region indicated below in Fig. ~\ref{sup fig: blue period}. The following plots reveal correlations between qubit frequency shifts and the voltage on nearby gates (there were not enough data points for qubits 3 and 4 in this period).

\begin{figure}[h]
    \centering
    \includegraphics[width=0.8\linewidth]{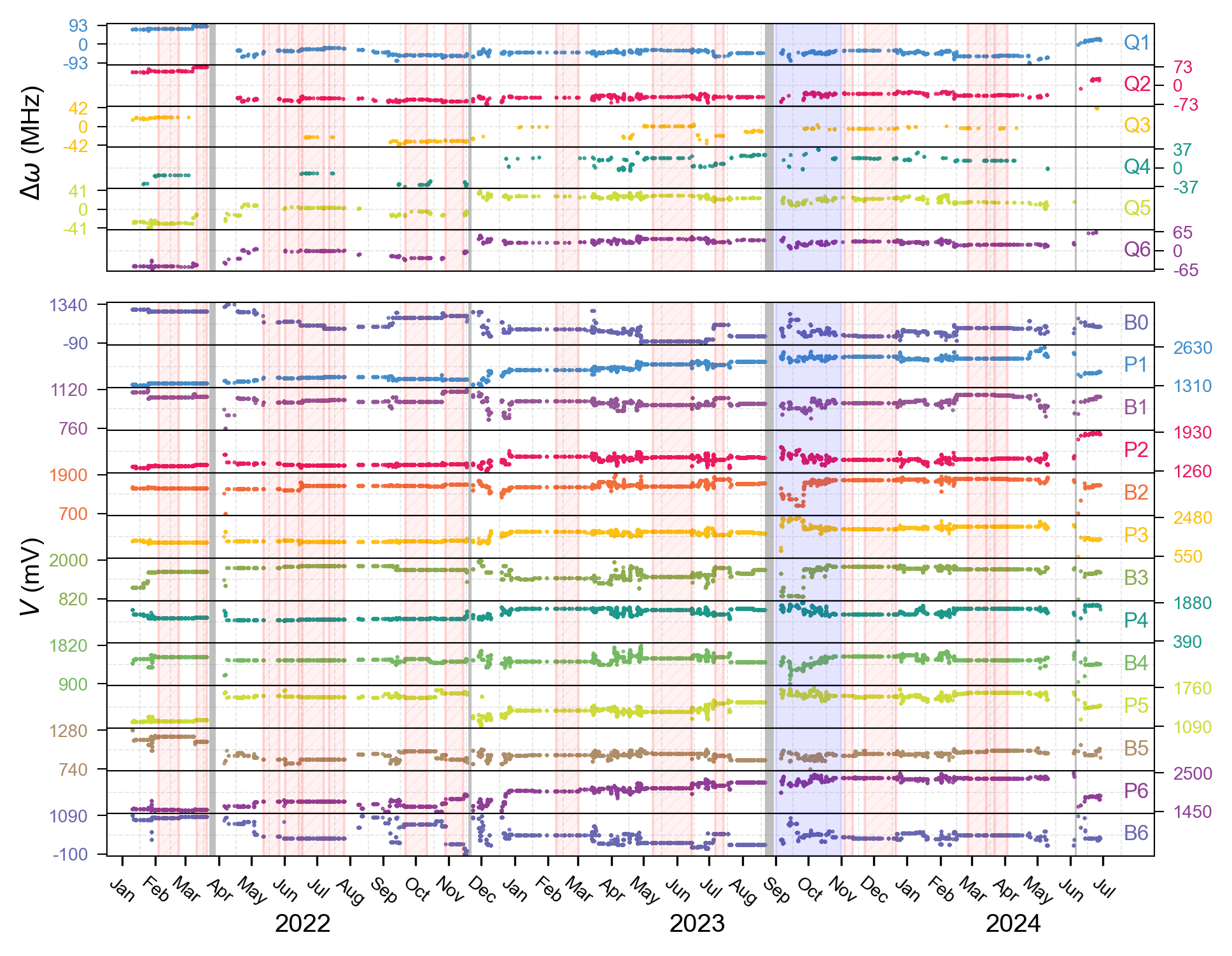}
    \caption{The blue region indicates the period of which we select the data from Fig. ~\ref{fig: frequency log}. This period was chosen because the gate voltages were varied over relatively large ranges, making it easier to detect possible correlations with qubit frequencies.}
    \label{sup fig: blue period}
\end{figure}

Correlations are not always present, even in some cases where they could have been expected. Possibly the effect of those gates on the corresponding qubit frequency was smaller than would be expected at first sight. Alternatively, since many gates were changed at a time, it is also quite possible that a shift in qubit frequency induced by a change in one gate voltage was cancelled by a shift in the voltage on another nearby gate. 

In some cases, correlations are observed where they are not expected, e.g. the gate voltage on B0 and P1 shows correlations with the frequency of qubit 6. It is plausible that in this case, we aimed to change the tunnel coupling between both outer qubits and the adjacent sensing dots (by adjusting B0 and compensating with P1 to keep the potential on dot 1 unchanged). This illustrates that we have to be cautious in drawing too strong conclusions from this analysis and that targeted experiments are needed to ascertain the correlation between individual gate voltages and qubit frequencies.

\begin{figure}[h]
    \centering
    \includegraphics[width=0.8\linewidth]{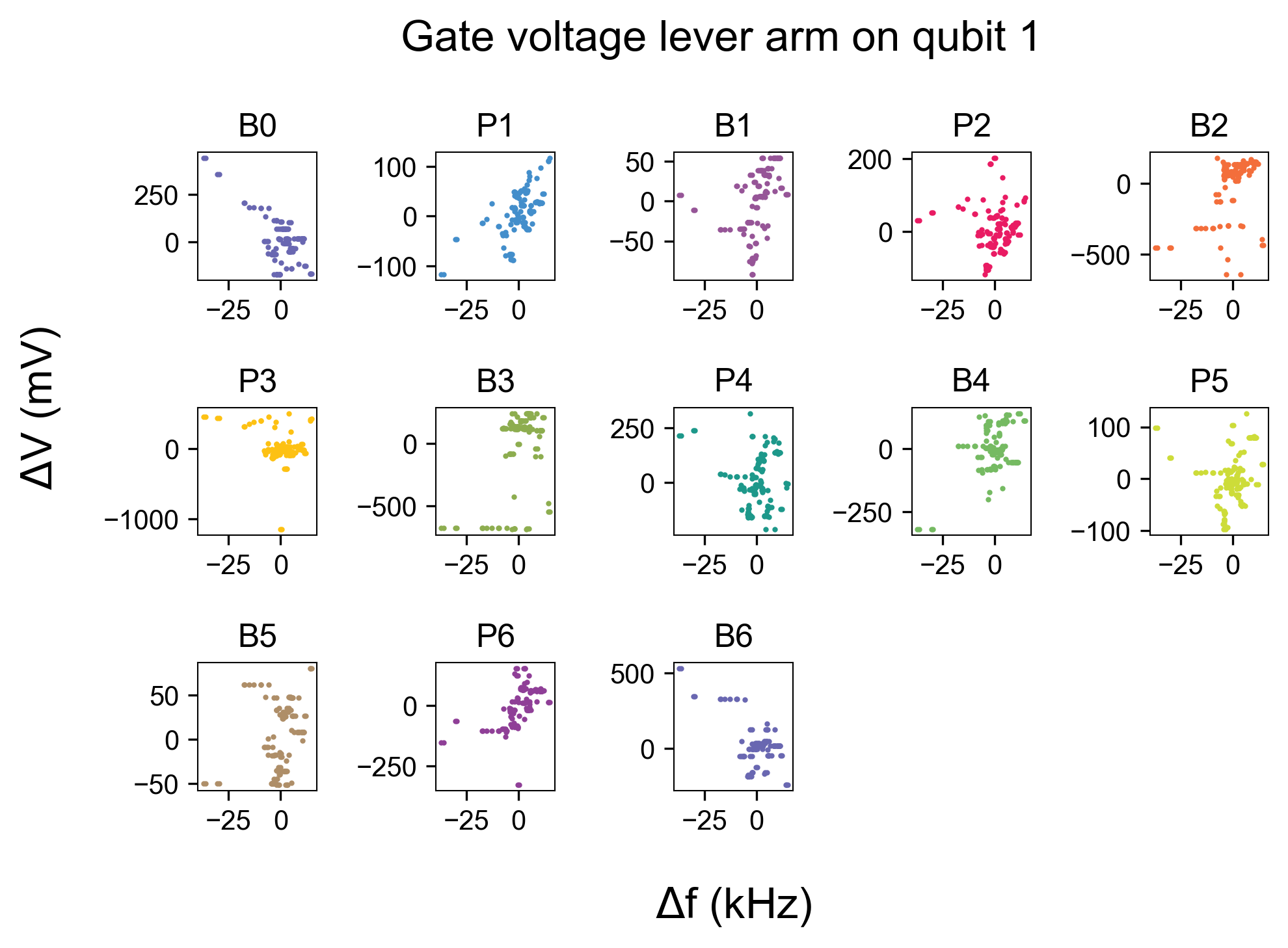}
    \caption{Qubit 1 frequency vs. gate voltage.}
    \label{sup fig: lever arms 1}
\end{figure}

\begin{figure}[h]
    \centering
    \includegraphics[width=0.8\linewidth]{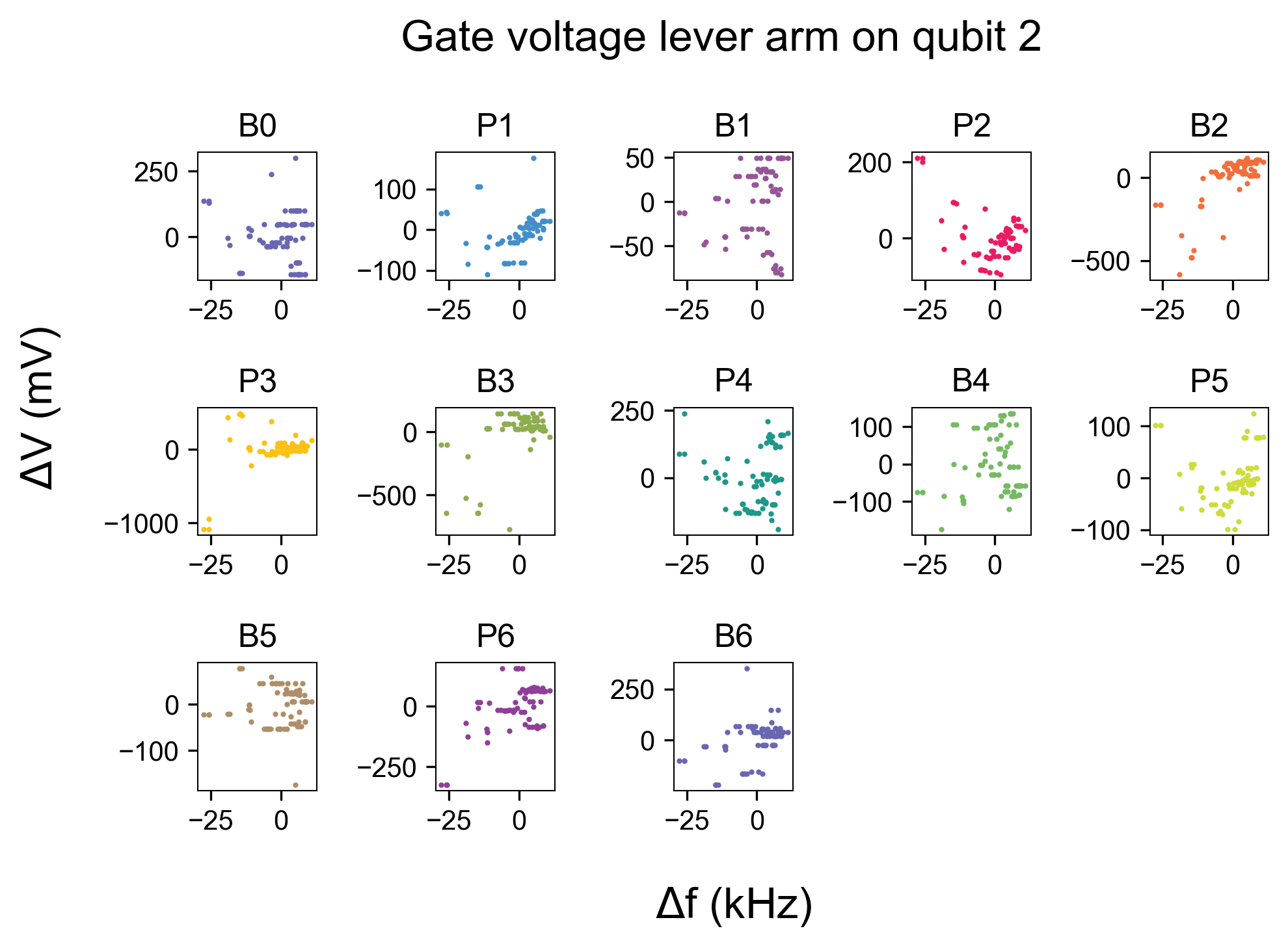}
    \caption{Qubit 2 frequency vs. gate voltage.}
    \label{sup fig: lever arms 2}
\end{figure}

\begin{figure}[h]
    \centering
    \includegraphics[width=0.8\linewidth]{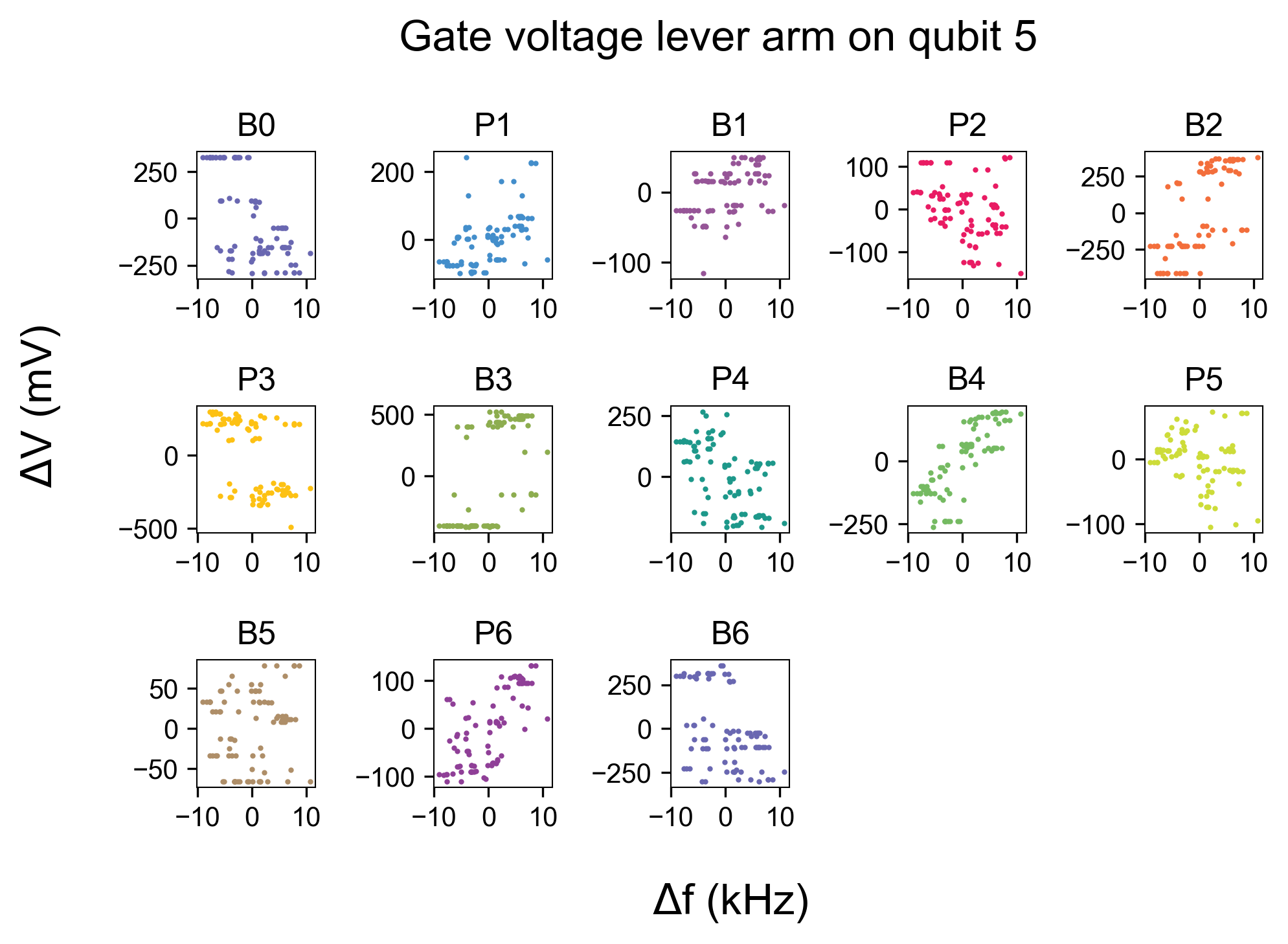}
    \caption{Qubit 5 frequency vs. gate voltage.}
    \label{sup fig: lever arms 5}
\end{figure}

\begin{figure}[h]
    \centering
    \includegraphics[width=0.8\linewidth]{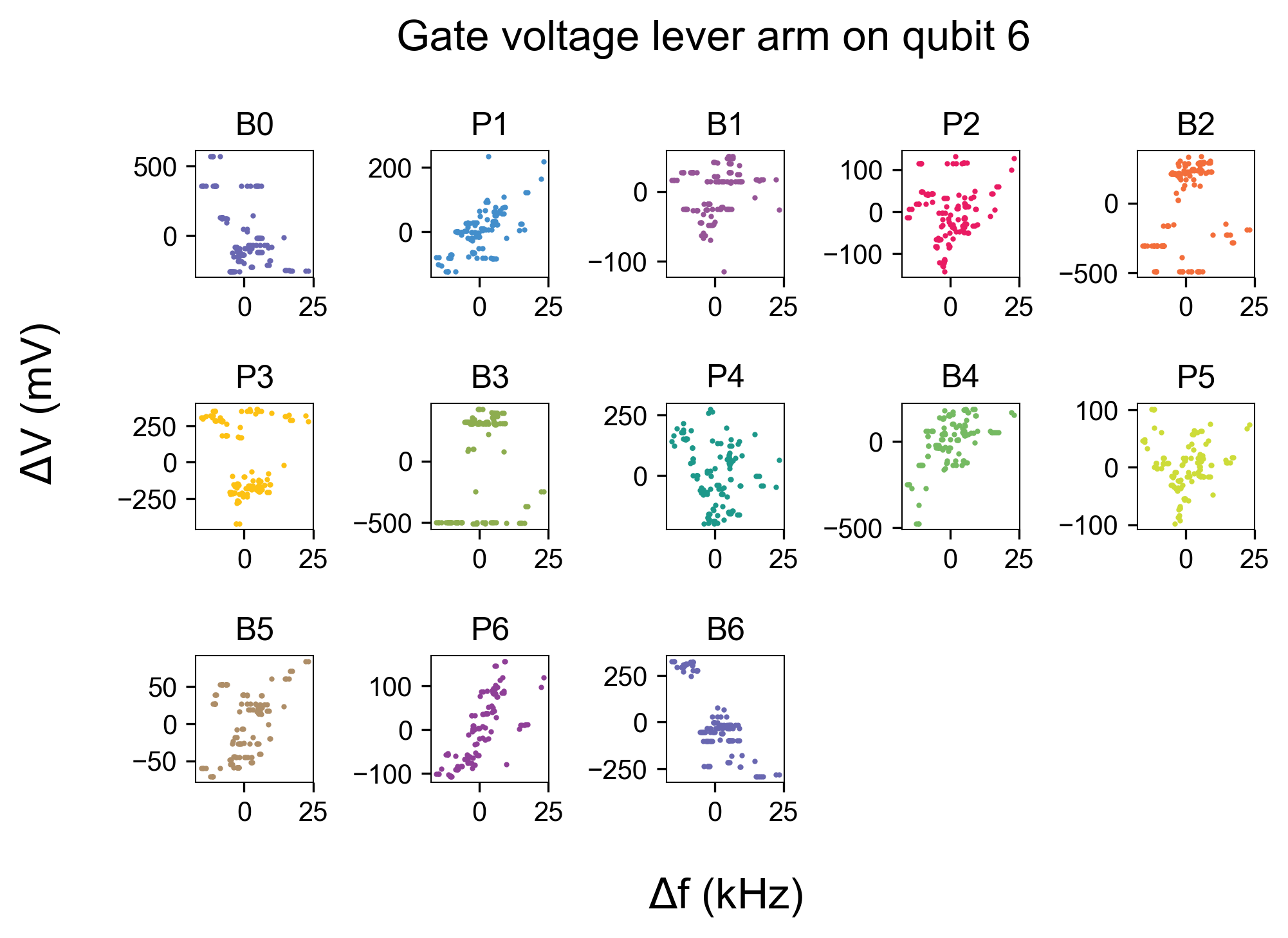}
    \caption{Qubit 6 frequency vs. gate voltage.}
    \label{sup fig: lever arms 6}
\end{figure}

\clearpage

\section{Qubit frequency drift data extracted from Ramsey experiments}\label{sup: drift data}
\subsection{Linear device}
\begin{figure}[ht]
    \centering
    \includegraphics[width=0.74\textwidth]{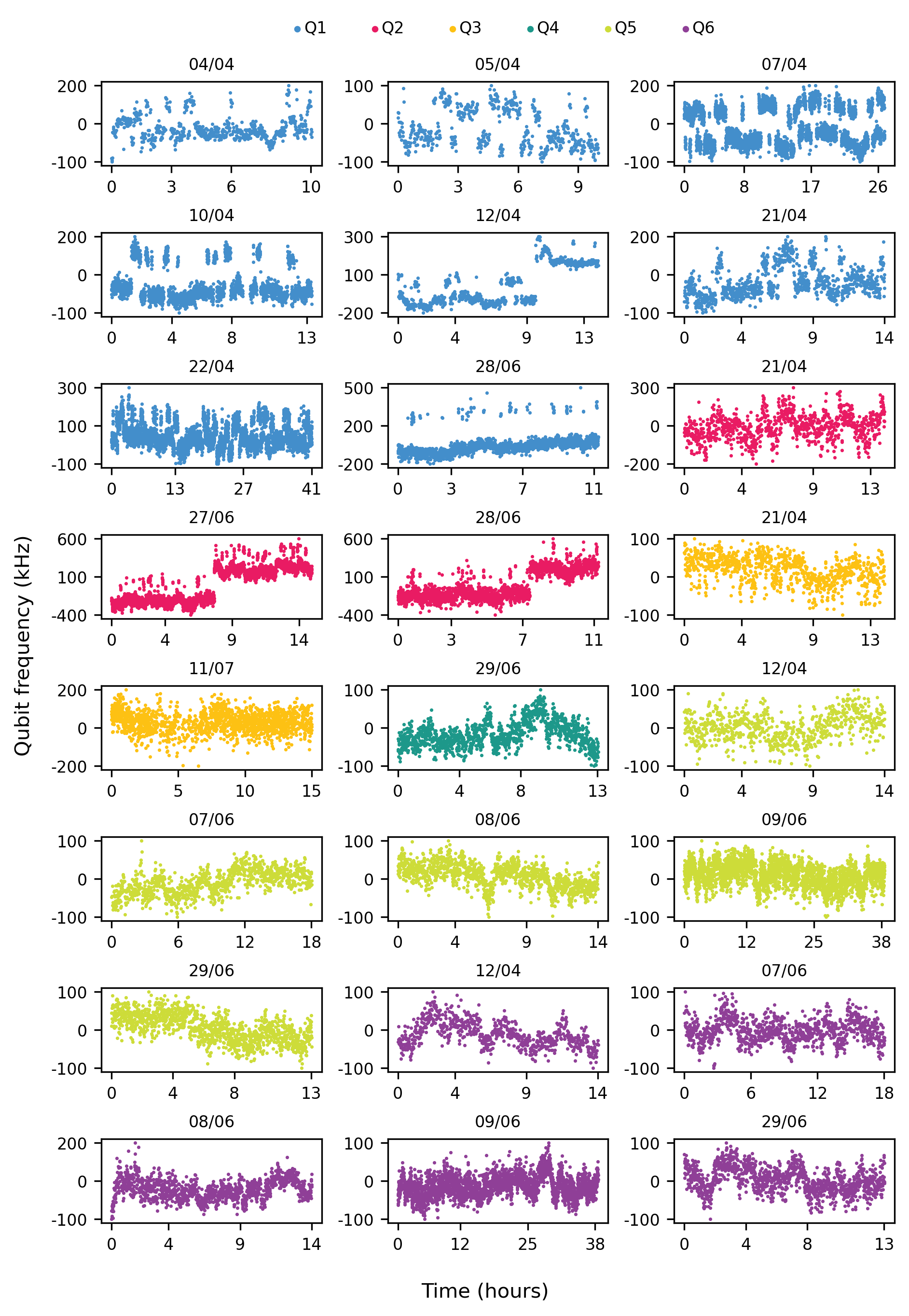}
    \caption{All obtained qubit frequency drift datasets based on Ramsey experiments from the linear device. The 24 figures are arranged in a 3x8 grid, ordered first by qubit number (1-6) and then for each qubit chronologically by measurement date. The arrangement follows a left-to-right, top-to-bottom order. All data was collected in 2023.}
    \label{sup fig: all drift 6D2S}
\end{figure}
\clearpage 

\subsection{2x2 device}
\begin{figure}[ht]
    \centering
    \includegraphics[width=0.74\textwidth]{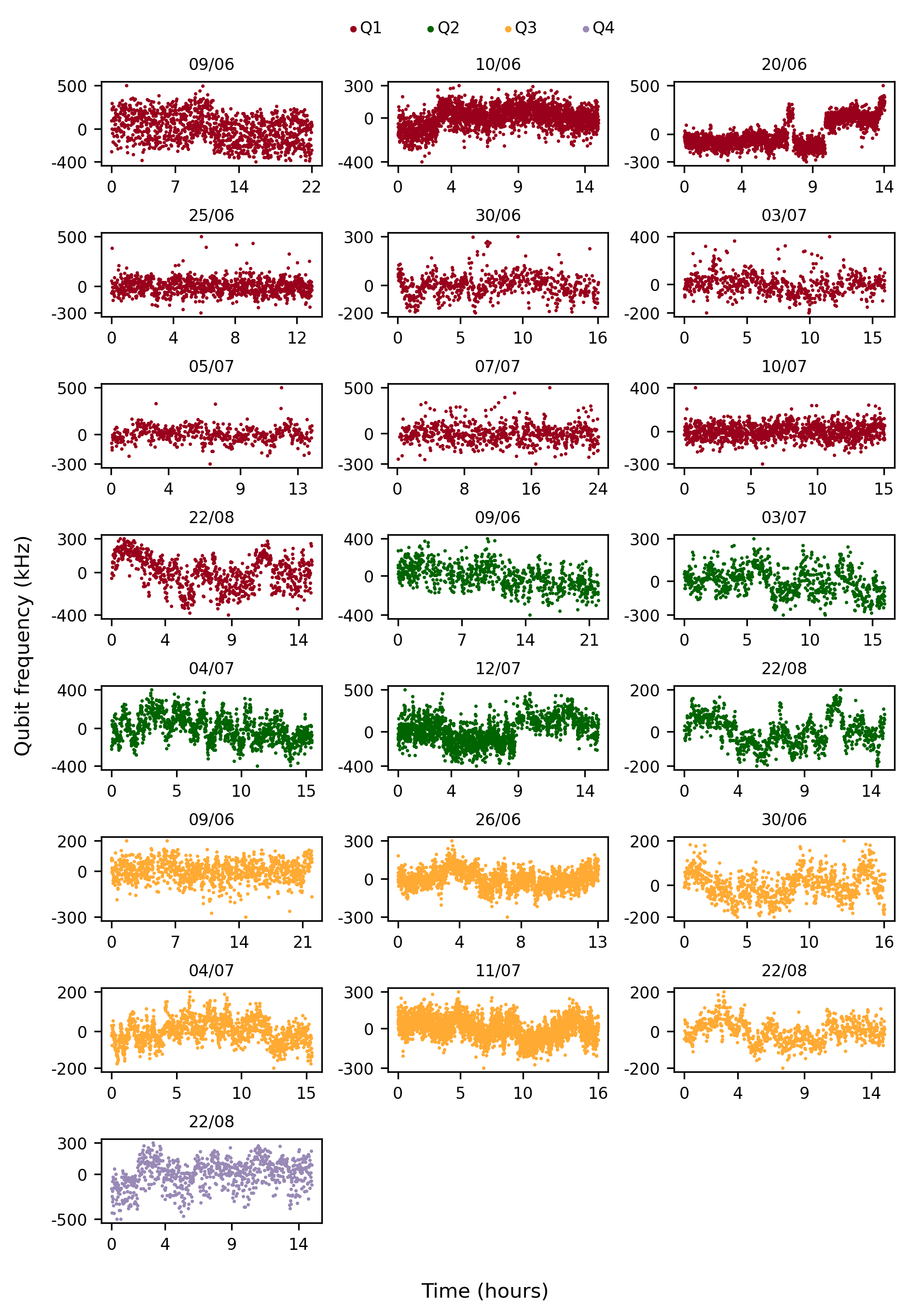}
    \caption{All obtained qubit frequency drift datasets based on Ramsey experiments from the 2x2 device. The 22 figures are arranged in a 3x8 grid, ordered first by qubit number (1-4) and then for each qubit chronologically by measurement date. The arrangement follows a left-to-right, top-to-bottom order. All data was collected in 2023.}
    \label{sup fig: all drift 2x2]}
\end{figure}
\clearpage 
\subsection{Information table for datasets}\label{sup: data table}
As described in the main text, each qubit frequency drift dataset consisted of Ramsey-style experiments which were repeated overnight. After extraction of the experiment data, we would fit each dataset and perform a round of filtering. We utilise the normalised chi-squared metric as described above in section~\ref{sup: filtering}. For all qubit frequency drift experiments filter out all datasets with a normalised chi-squared score of $\tilde{\chi}^2\geq0.8$.\\

Below we present tables containing information on the datasets collected from both devices. The columns indicate, in order:
\begin{enumerate}
    \item Date of data collection
    \item Qubit number
    \item Total data collection time
    \item Total number of oscillations
    \item Number of filtered datapoints (i.e. datasets with a normalised chi-squared score of 0.8 or above)
    \item Percentage of filtered datapoints with respect to the total number of datapoints
    \item Average frequency error ($1\sigma$) extracted from the covariance matrix of the fit
    \item Average normalised chi-squared metric of all datapoints after the filtering
\end{enumerate}

The red bars correspond to datasets that required a large amount of filtering (more than 5\% of total datapoints). These datasets were included for the average 1-hour frequency deviation results in Fig.~\ref{fig: freq drift deviation} of the main text, however were deemed unsuitable for Fourier analysis in Fig.~\ref{fig: PSDs}, due to the requirement of the data being sampled uniformly in time. 

\begin{figure}[ht]
    \centering
    \includegraphics[width=\linewidth]{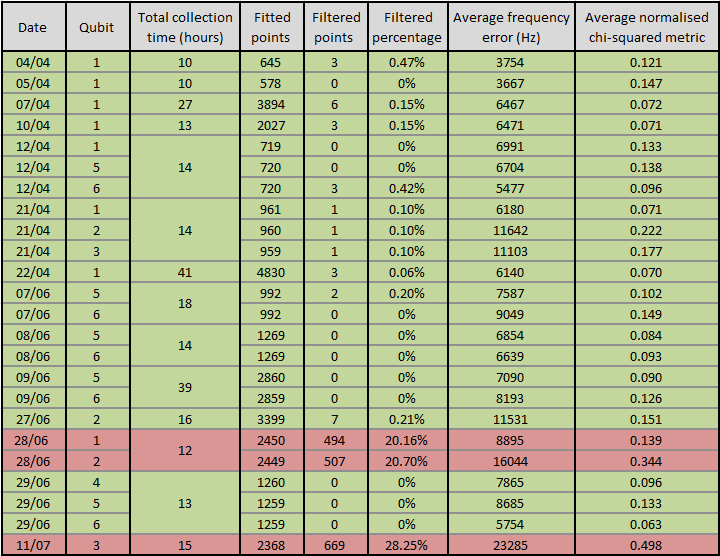}
    \caption{Data table for all drift datasets obtained in the linear device.}
    \label{sup fig: data table 6D2S}
\end{figure}
\begin{figure}[ht]
    \centering
    \includegraphics[width=\linewidth]{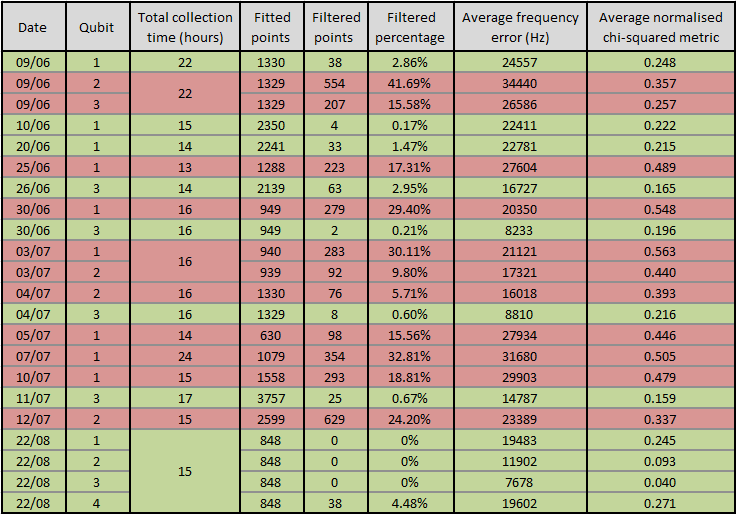}
    \caption{Data table for all drift datasets obtained in the 2x2 device.}
    \label{sup fig: data table 2x2]}
\end{figure}
\clearpage 
\subsection{DC voltage logs during Ramsey experiments}\label{sup: DC logs}
In this section we show the DC voltage settings of both devices during the Ramsey-style experiments. Changes in these settings are likely to affect the qubit frequency noise that each qubit experiences. The gates are labeled as according to Fig.~\ref{fig: devices} in the main text. All Ramsey experiments for both the linear and 2x2 devices were collected in 2023. 

\begin{figure}[ht]
    \centering
    \includegraphics[width=\textwidth]{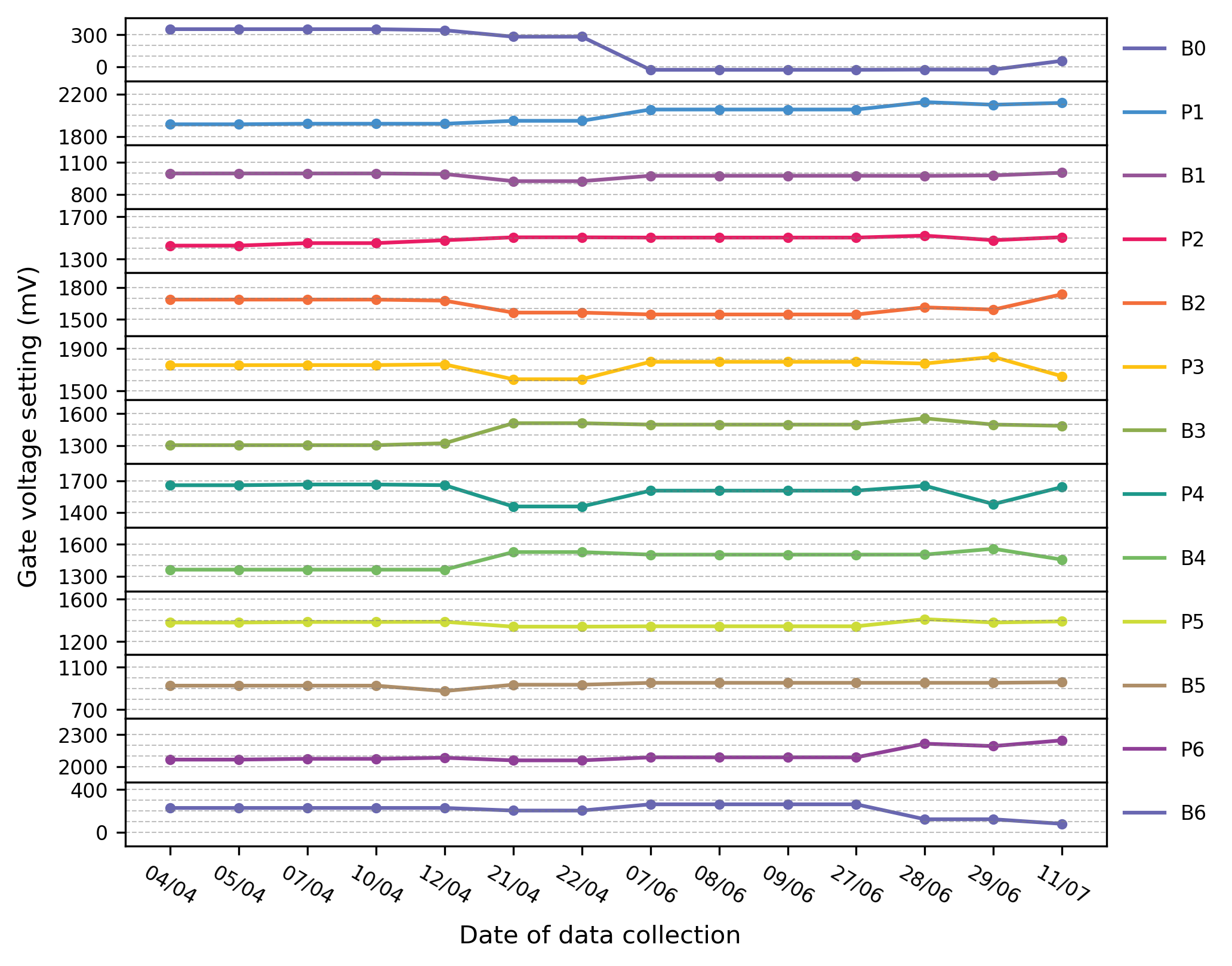}
    \caption{DC voltage settings in the linear device. In each subplot, dashed horizontal lines are plotted at multiples of 100 mV. The mixing chamber temperature throughout the data collection process were as follows: (i) 04/04 - 22/04, 200 mK, (ii) 07/06 - 09/06 300 mK and (iii) 27/06 - 11/07, 200mK.}
    \label{sup fig: DC voltages 6D2S}
\end{figure}
\clearpage
\begin{figure}[ht]
    \centering
    \includegraphics[width=0.5\textwidth]{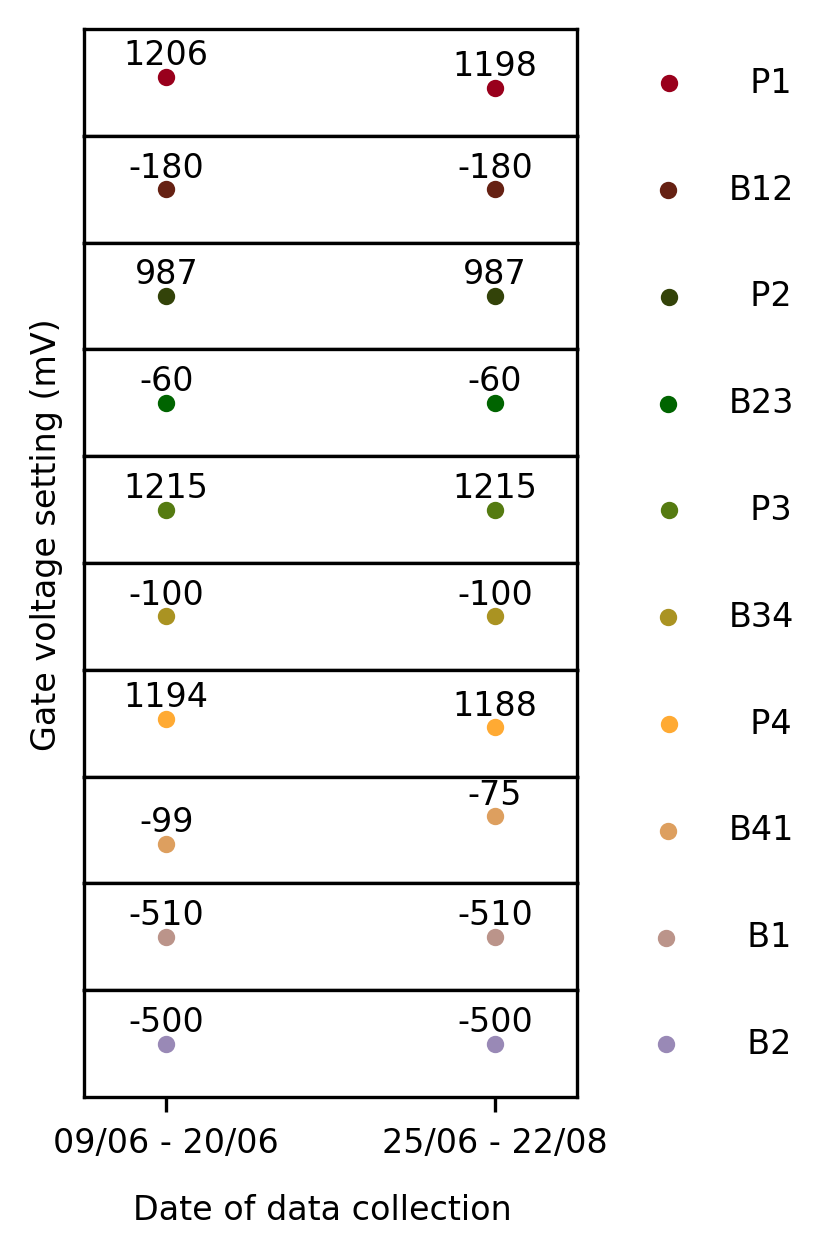}
    \caption{DC voltage settings in the 2x2 device. Throughout the Ramsey data collection period, this device was only operated in two distinct voltage configurations, between 09 - 20/06 and 25/06 - 22/08. The only three gates to undergo a change are P1, P4 and B41. The mixing chamber temperature throughout the data collection process were as follows: (i) 09/06 - 10/06, 11 mK (ii) 20/06 - 12/07, 200 mK and (iii) 22/08, 11 mK.}
    \label{sup fig: DC voltages 2x2}
\end{figure}
\clearpage 
\section{Strong TLF and discrete jump events affecting qubit frequencies}\label{sup: TLF and discrete jumps}
\subsection{Characterisation of TLF coupling to qubit 1}
From the 24 collected 10-41 hour Ramsey datasets in the linear device, we can often observe a two-level fluctuator (TLF) with a strong coupling to the qubit frequency of qubit 1. By fitting the histograms of these time traces to a double gaussian, we can extract the induced frequency shift by this TLF. 
\begin{figure}[ht]
    \centering
    \includegraphics[width=\textwidth]{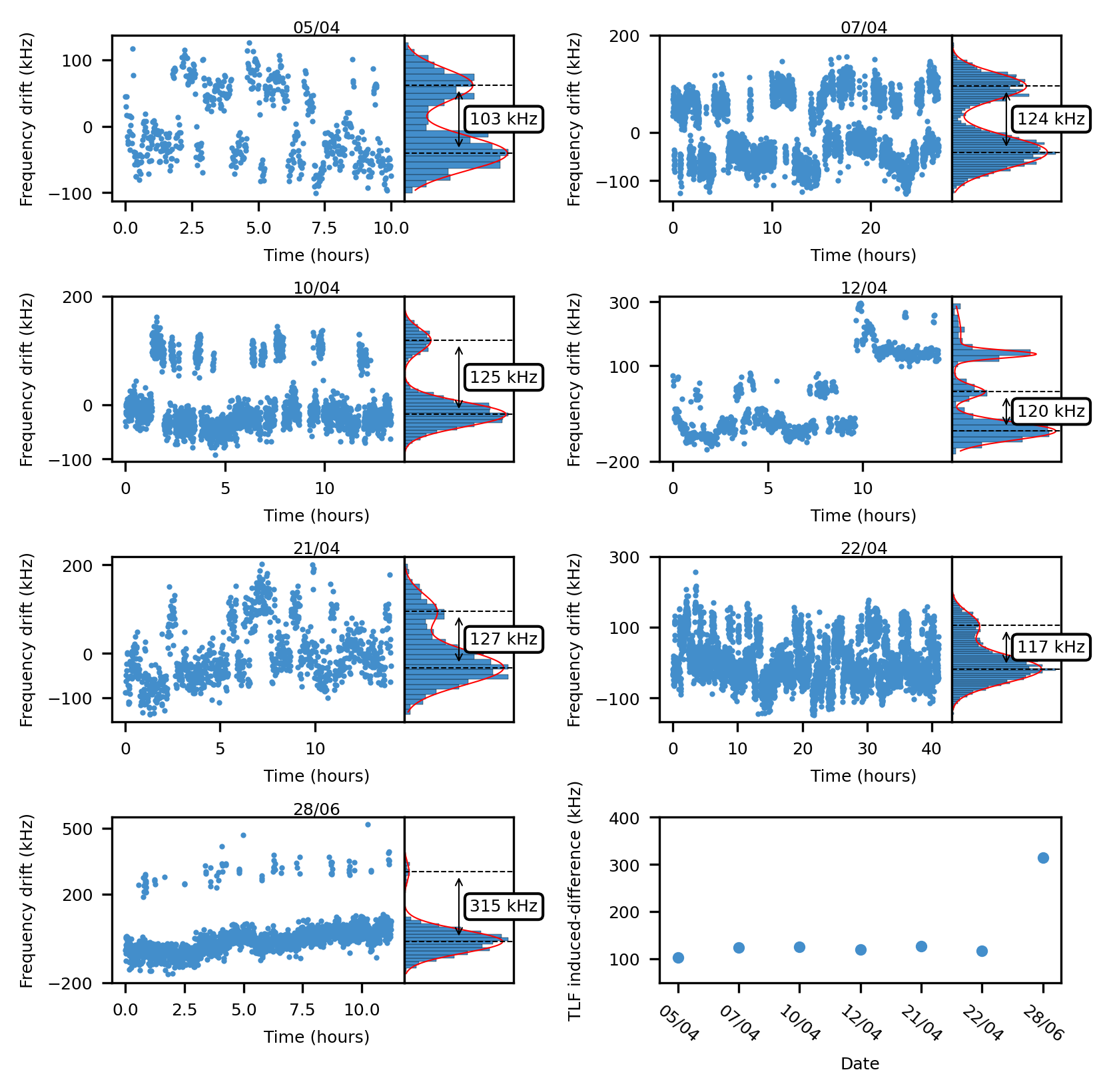}
    \caption{Different time traces capturing the behaviour of the TLF coupling to qubit 1, from 05/04 to 22/08. In the final subplot (bottom-right), we show the qubit frequency shift induced by the TLF on each date of data collection.}
    \label{sup fig: TLF behaviour}
\end{figure}

From the final plot, we can see how the behaviour of the TLF changes. The induced frequency shift is approximately about 120 kHz from 05/04 - 22/04. However, on 28/06, we can see a much larger induced difference, of about 315 kHz. The strength of the coupling of the TLF to the qubit is likely dependent on the gate voltage settings. 

\subsection{Discrete jump events}
Here we show three instances of large discrete jumps. Whilst these events induced large frequency shifts (of hundreds of kHz), the occurrences are quite rare having only been observed on three occasions in the course of the 24 Ramsey measurements on the linear device. On 28/06, we interleaved qubit 1 and 2 frequency drift measurements. We observe a large discrete jump in the frequency of qubit 2 but not in qubit 1, thus we speculate that the origin of this particular jump is sufficiently localized to the second dot.

\begin{figure}[ht]
    \centering
    \includegraphics[width=\textwidth]{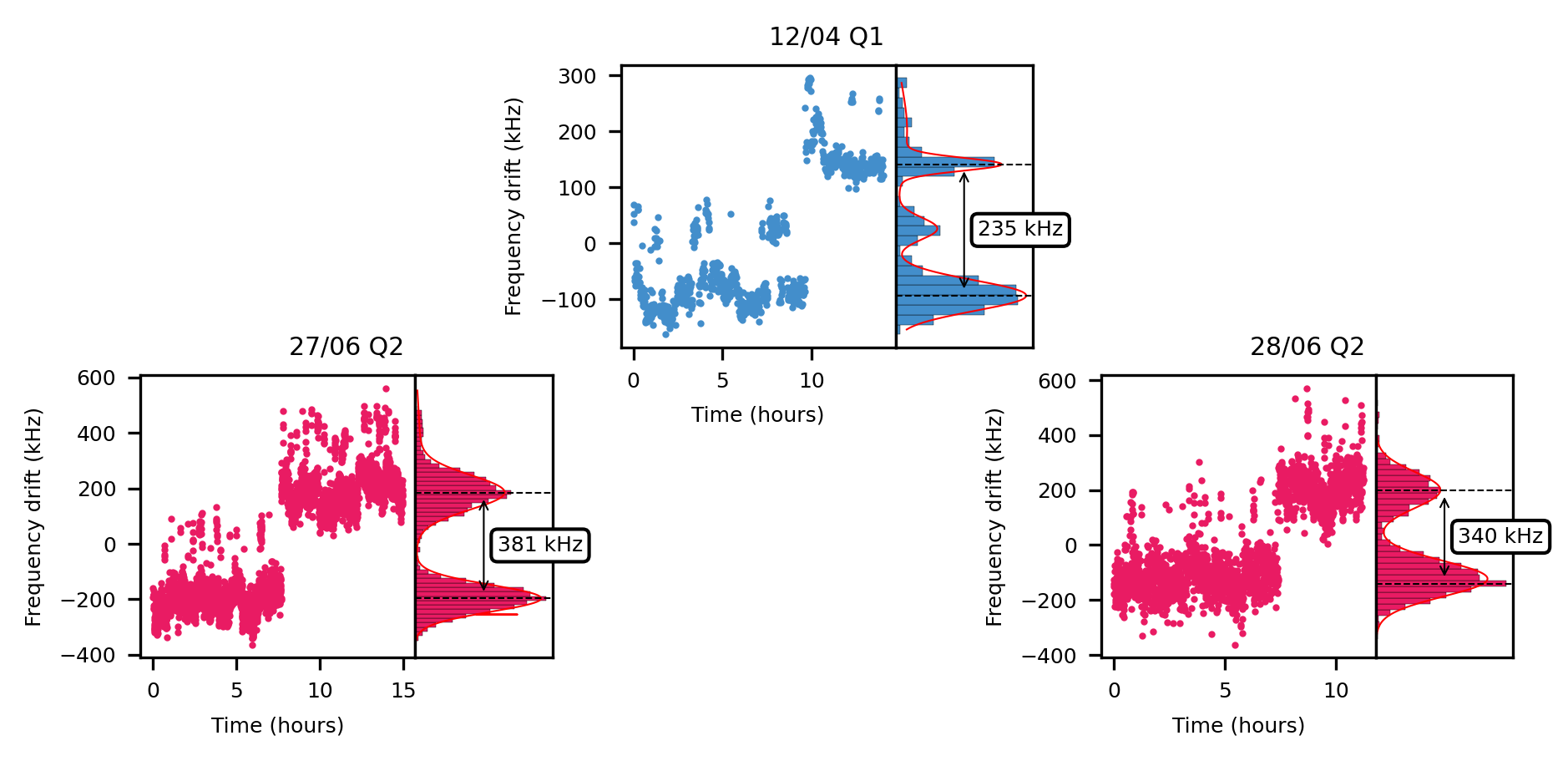}
    \caption{We show three recorded events of a large discrete jump in qubit frequency extracted from Ramsey experiments in the linear device, for qubit 1 on 12/04 of magnitude 235 kHz and qubit 2 on 27/06 of magnitude 381 kHz and 28/06 of magnitude 340 kHz.}
    \label{sup fig: discrete jumps}
\end{figure}

\newpage 
\section{Gate fidelity of $X_{\pi/2}$ gate under coherent qubit frequency error}\label{sup: gate fidelity}
To understand the consequences that qubit frequency detuning has on single-qubit gate fidelities, we consider the EDSR Hamiltonian from Eq.~\ref{eq: EDSR rotating frame} from the main text and use this to derive the gate fidelity of the $X_{\pi/2}$ gate as a function of qubit frequency detuning. The choice of this gate in particular is arbitrary, and the following derivation can also be repeated for a different rotation axis in the $X-Y$ plane. We use the equation for the average gate fidelity as provided in~\cite{pedersen2007fidelity}:

\begin{equation}\label{Average gate fidelity}
    F = \frac{d+|\text{Tr}[U_{ideal}^\dag U_{actual}]|^2}{d(d+1)}
\end{equation}

where $U_{ideal}$ corresponds to the ideal operation and $U_{actual}$ the physical realisation of this operation. The variable $d$ is the dimension of the Hilbert space in which the operations act, thus $d=2$ for single-qubit operations. We assume that the implementation of the gate is perfect aside from the effects of qubit frequency detuning. We first rewrite the EDSR Hamiltonian from Eq.~\ref{eq: EDSR rotating frame} as 

\begin{equation}\label{Drive Hamiltonian 2}
    H = \frac{\Delta}{2}\sigma_z + \frac{\Omega}{2}\sigma_x = \frac{\lambda}{2}\vec{n}\cdot\vec{\sigma}
\end{equation}
where $\lambda = \sqrt{\Delta^2+\Omega^2}$, $\vec{n} = \frac{1}{\lambda}(\Omega, 0, \Delta)$ and $\vec{\sigma} = (\sigma_x, \sigma_y, \sigma_z)$. We further make the assumption that the detuning $\Delta$ is quasi-static. Using this Hamiltonian we can simulate the implementation of an EDSR burst for a duration $t$ with detuning $\Delta$ as a unitary evolution, corresponding to time-evolution under the Hamiltonian $H$:

\begin{equation}\label{U actual}
\begin{aligned}
    U_{actual}(t, \Delta) &= e^{-iHt} = e^{-i\vec{n}\cdot\vec{\sigma}\lambda t/2}\\
    &= \text{cos}(\lambda t/2)\mathbb{I} -i\text{sin}(\lambda t/2)\vec{n}\cdot\vec{\sigma}\\
    &= \begin{pmatrix}
        \text{cos}(\lambda t/2) -i\frac{\Delta}{\lambda}\text{sin}(\lambda t/2) & -i\frac{\Omega}{\lambda}\text{sin}(\lambda t/2)\\
        -i\frac{\Omega}{\lambda}\text{sin}(\lambda t/2) & \text{cos}(\lambda t/2) +i\frac{\Delta}{\lambda}\text{sin}(\lambda t/2)
        \end{pmatrix}
\end{aligned}
\end{equation}

For zero-detuning ($\Delta=0\implies\lambda=\Omega$) and fixing a pulse duration of $t_{\pi/2}=\frac{\pi}{2\Omega}$, we obtain the ideal $X_{\pi/2}$ rotation

\begin{equation}
    U_{actual}(t=t_{\pi/2}, \Delta=0) = U_{ideal} = \frac{1}{\sqrt{2}}\begin{pmatrix}1&-i\\-i&1\end{pmatrix}
\end{equation}

Using this expression, we can see that

\begin{equation*}
\begin{aligned}
    \text{Tr}[U_{ideal}^\dag U_{actual}(t,\Delta)] &= \frac{1}{\sqrt{2}}\Bigl[2\text{cos}(\lambda')+2\frac{\Omega}{\lambda}\text{sin}(\lambda')\Bigr]\\
    \implies |\text{Tr}[U_{ideal}^\dag U_{actual}(t,\Delta)]|^2
    &= 2\Bigl[\frac{\Omega^2}{\lambda^2}+\frac{\Delta^2}{\lambda^2}\text{cos}^2(\lambda')+\frac{\Omega}{\lambda}\text{sin}(2\lambda')\Bigr]
\end{aligned}
\end{equation*}
where $\lambda' = \frac{\lambda\pi}{4\Omega}$. Therefore plugging this expression back into Eq.~\ref{Average gate fidelity} we get the fidelity of the $X_{\pi/2}$ gate as a function of the detuning $\Delta$ and the Rabi frequency $\Omega$

\begin{equation}\label{Fidelity vs. detuning}
    F(\Delta, \Omega) = \frac{1}{3}\Bigl[1+\frac{\Omega^2}{\lambda^2}+\frac{\Delta^2}{\lambda^2}\text{cos}^2\bigl(\frac{\lambda\pi}{4\Omega}\bigr)+\frac{\Omega}{\lambda}\text{sin}\bigl(\frac{\lambda\pi}{2\Omega}\bigr)\Bigr]
\end{equation}

\newpage 
\section{Power Spectral Densities}\label{sup: PSDs}
Here we show all obtained power spectral densities (PSDs) from the time domain data collected via the Ramsey experiments. The dashed red lines are fits to the equation $S(f) = S_0/f^\alpha$, and the fitted parameters $S_0$ and $\alpha$ for all datasets are plotted in Fig.~\ref{fig: PSDs}b of the main text.

\subsection{Linear device}
\begin{figure}[ht]
    \centering
    \includegraphics[width=0.67\textwidth]{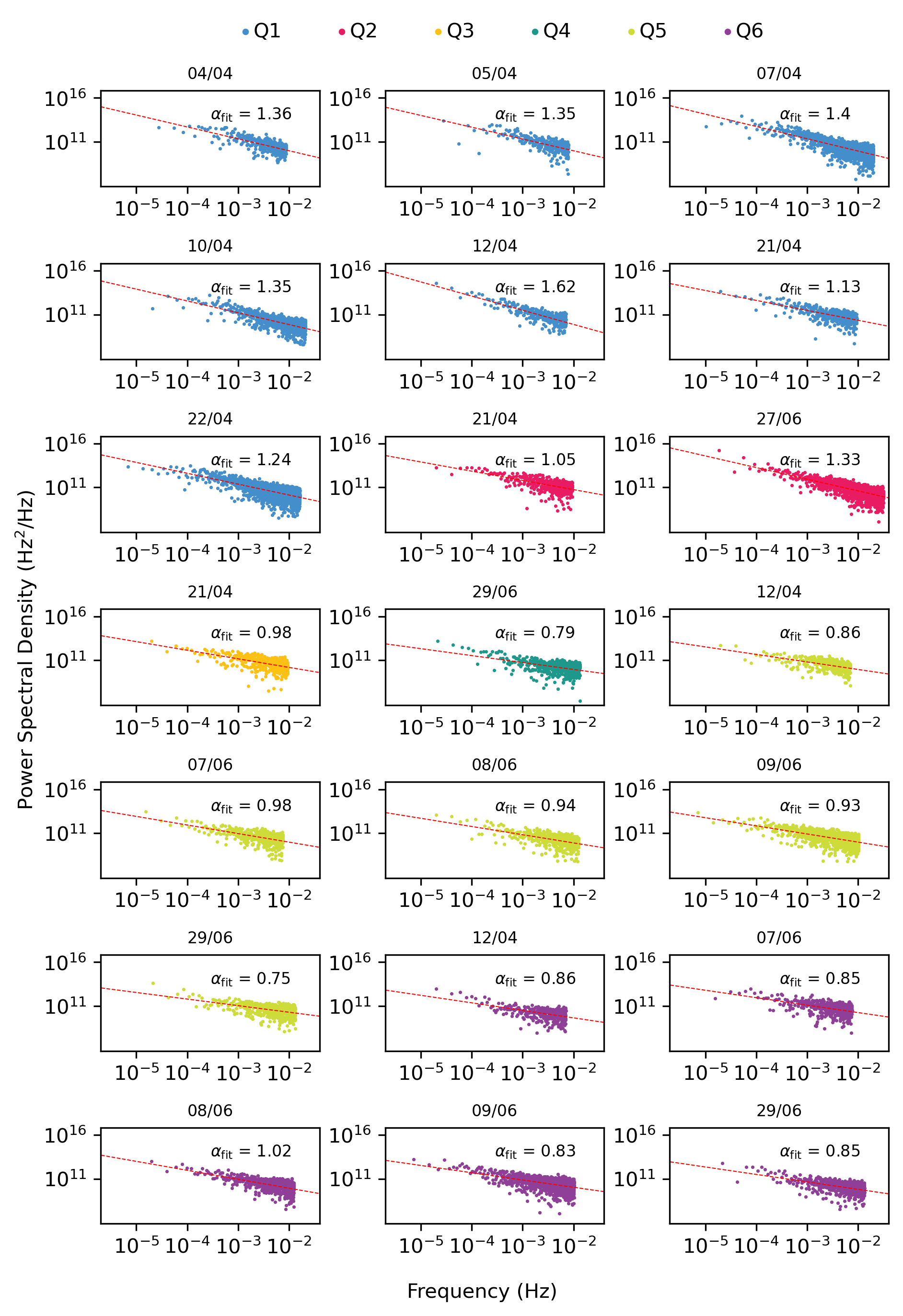}
    \caption{All obtained PSDs from the linear device. The 21 figures are
arranged in a 3x8 grid, ordered first by qubit number (1-6) and then for each qubit chronologically by measurement date. The arrangement follows a left-to-right, top-to-bottom order. All data was collected in 2023}
    \label{fig:enter-label}
\end{figure}
\newpage 

\subsection{2x2 device}
\begin{figure}[ht]
    \centering
    \includegraphics[width=0.9\textwidth]{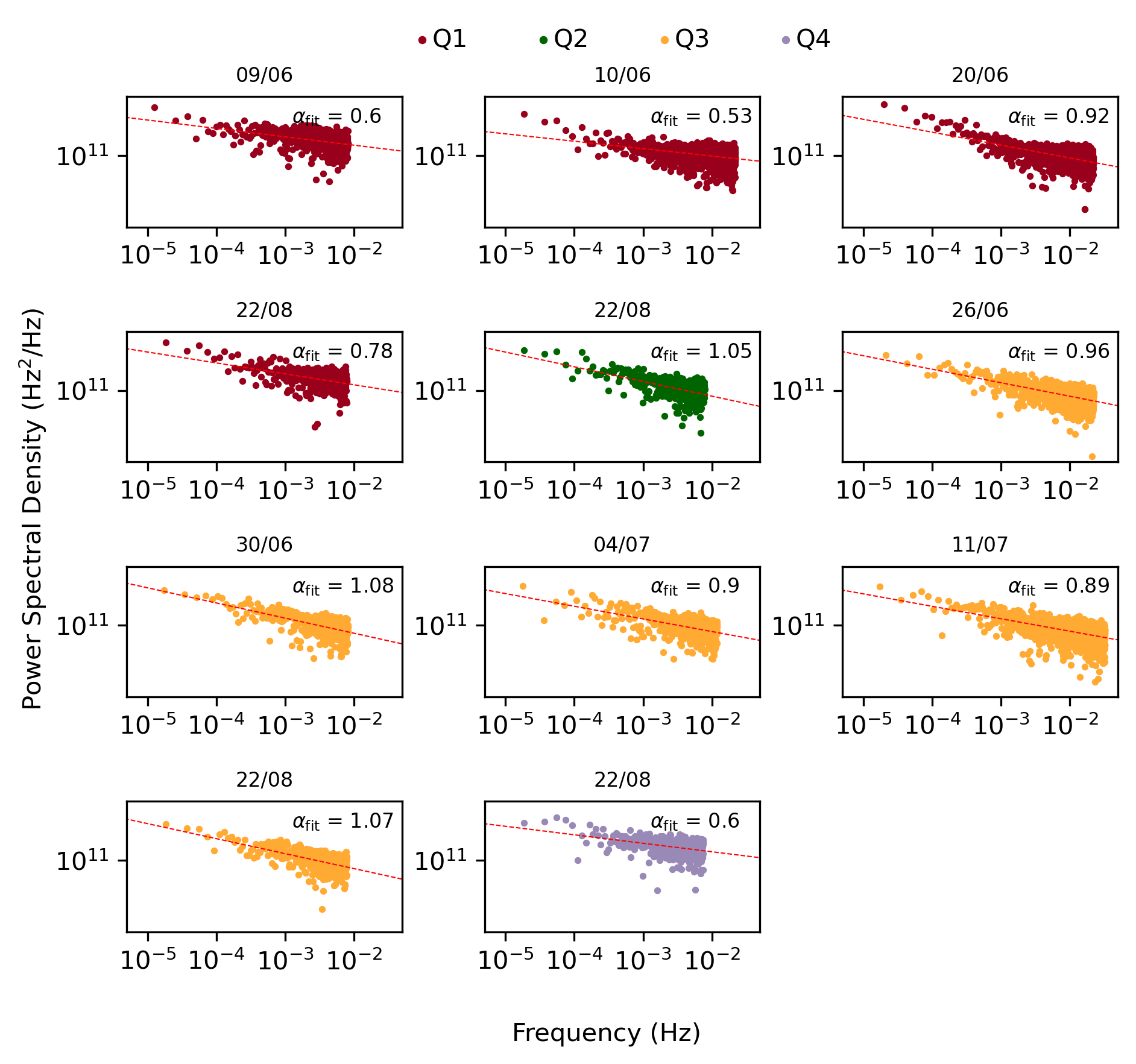}
    \caption{All obtained PSDs from the 2x2 device. The 11 figures are
arranged in a 3x8 grid, ordered first by qubit number (1-4) and then for each qubit chronologically by measurement date. The arrangement follows a left-to-right, top-to-bottom order. All data was collected in 2023}
    \label{fig:enter-label}
\end{figure}

\subsection{PSD from frequency calibration data}\label{sup: freq calib PSD}

The frequency calibration data suffers from both a low number of datapoints and a non-uniform sampling rate, as we are utilising data from calibration routines that were sporadically collected and not initially intended for this analysis. To overcome this, we first implement a cubic spline interpolation to transform the data onto a uniform time axis, followed by the regular Fourier analysis on the transformed dataset. This is the method we use to obtain the PSDs presented in Fig.~\ref{fig: PSDs}a of the main text.

There also exists the Lomb-Scargle method to extract the PSD from data that is non-uniformly sampled in the time domain. We test this method, along with the aforementioned method (which we call the ``interpolation'' method), on artificially generated data with a $1/f$ noise spectrum. We generate this data by first creating an artificial time trace with steps of $0.0001$s for a total duration of $100$s (thus this signal has in total $N=1,000,000$ points). At each time-step, we simulate the effect of 100 randomly generated TLFs. These TLFs are generated with a normally distributed strength $s_i$ (i.e. each individual TLF modulates the signal by $\pm s_i$) with mean $\mu_s=0.1$ and standard deviation $\mu_\sigma = 0.1$ and a log-uniformly distributed switching rate $\tau_i$ (i.e. the average rate at which it switches between the two configurations $[-s_i, +s_i]$) sampled between $\tau_{min} = 10^{-4}$Hz and $\tau_{max} = 10^4$Hz. Finally, at each time step, we also add a normally distributed noise term with mean $\mu=0$ and standard deviation $\sigma=0.01$ to capture the effects of measurement errors. In Fig.~\ref{sup fig: fake TLF data} below, we can see an example of the artificially generated TLF data in both the time and frequency domain. 

\begin{figure}[ht]
    \centering
    \includegraphics[width=\textwidth]{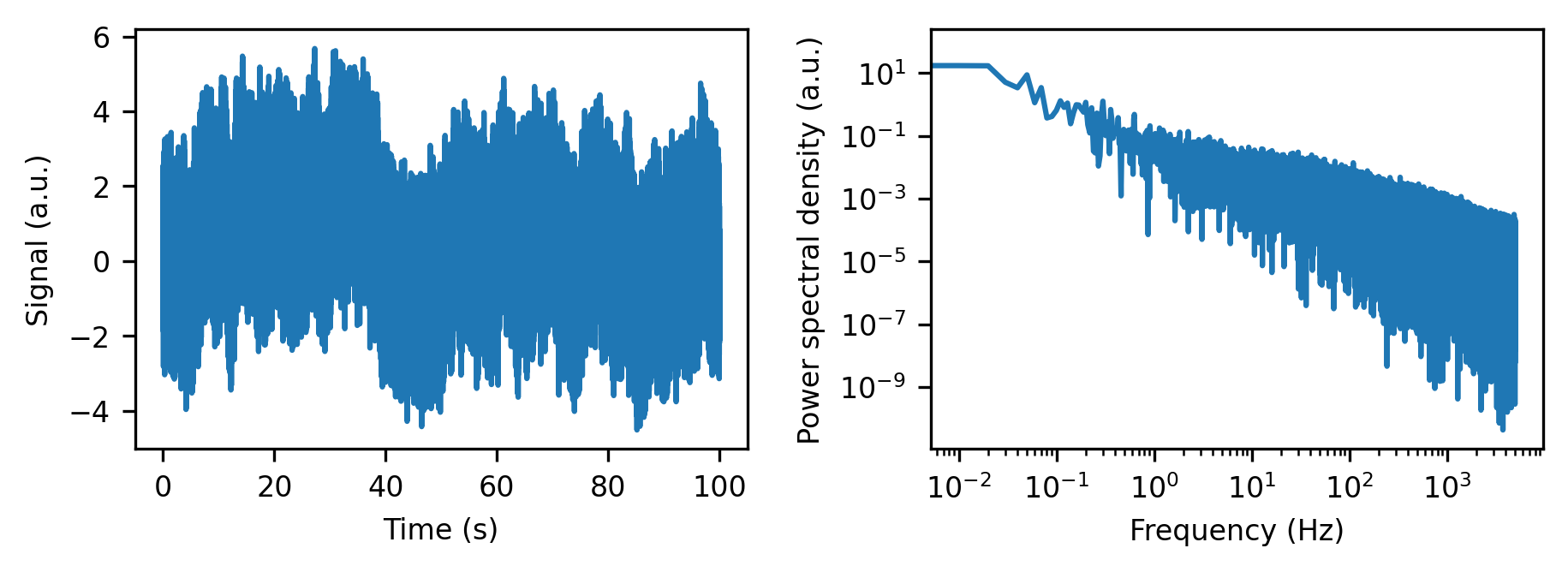}
    \caption{Artificially generated time- and frequency-domain data, simulating the effect of 100 TLFs.}
    \label{sup fig: fake TLF data}
\end{figure}

To test both the interpolation and Lomb-Scargle methods, we first sample from the time-domain dataset. We ensure that the sampled signal is based on a non-uniform sampling rate and consists of only 25 datapoints. We then apply both methods on the sampled signal and compare the obtained PSDs with the PSD of the full signal. This is presented below in Fig.~\ref{sup fig: compare methods PSD}. We can observe that both methods reproduce the original PSD with reasonable accuracy in their respective frequency range, with in this case less scatter for the interpolation method. We performed a number of similar tests using artificially generated noise to further validate these methods. 
\begin{figure}[ht]
    \centering
    \includegraphics[width=\textwidth]{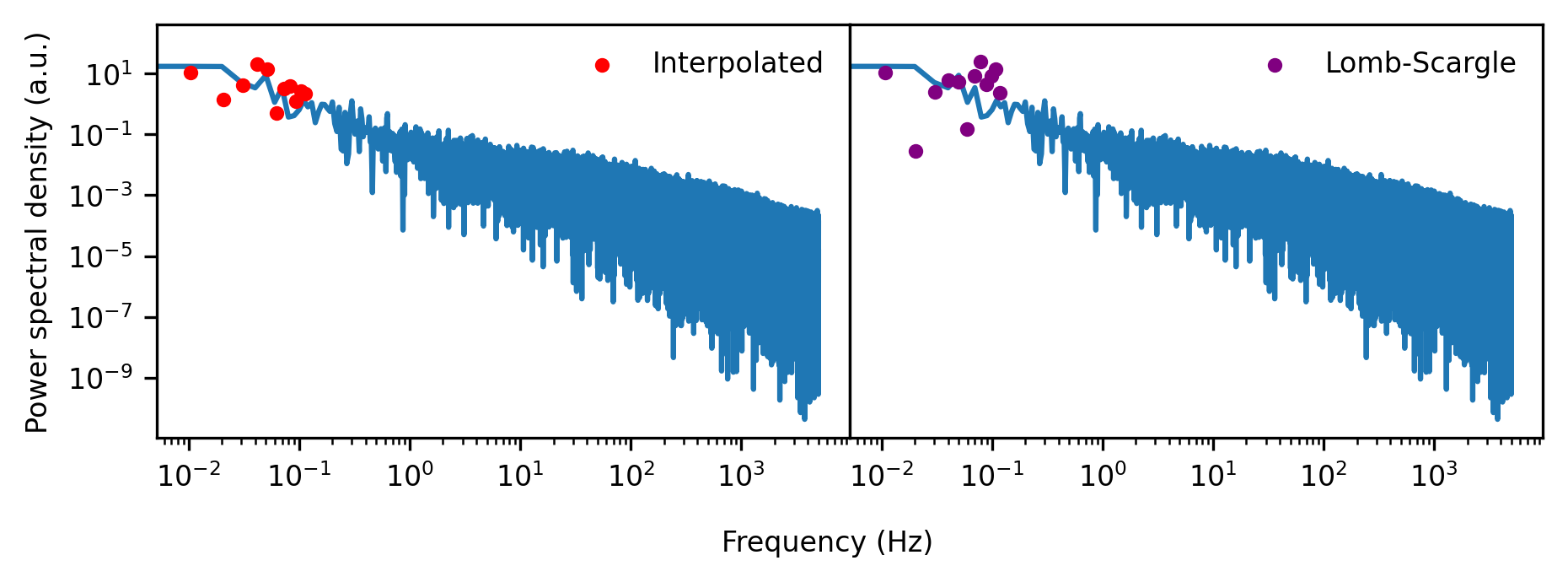}
    \caption{Comparing the output of the interpolation method (left, in red) vs. the Lomb-Scargle method (right, in purple) at recreating the original PSD for a non-uniformly sampled signal with only 25 datapoints.}
    \label{sup fig: compare methods PSD}
\end{figure}

\end{document}